\documentclass[11pt]{article}
\usepackage{graphics,amssymb}
\def\lm{\lambda}
\usepackage{epsfig}
 
\begin{document}
 
{
\setlength{\textwidth}{16.5cm}
\setlength{\textheight}{22.2cm}
\setlength{\hoffset}{-1.43cm}
\setlength{\voffset}{-.9in}

\thispagestyle{empty}
\renewcommand{\thefootnote}{\fnsymbol{footnote}}

\begin{flushright}
{\normalsize
SLAC-AP-134\\
LCC-0043\\
December 2000}
\end{flushright}

\vspace{.8cm}

\begin{center}
{\bf\Large Dipole Mode Detuning in the\break Injector Linacs of the NLC
\footnote{\small Work supported by
Department of Energy contract  DE--AC03--76SF00515.}}

\vspace{1cm}

{\large
Karl L.F. Bane and Zenghai Li\\
Stanford Linear Accelerator Center, Stanford University,
Stanford, CA  94309}

\medskip

\end{center}

\begin{center}
{\bf\large   
Abstract }
\end{center}

\begin{quote}
The injector linacs of the JLC/NLC
project include the prelinac, the $e^+$ drive linac,
the $e^-$ booster, and the $e^+$ booster. The first three will
be S-band machines, the last one, an L-band machine.
We have demonstrated that by using detuning alone in the 
accelerator structure design of these linacs
we will have acceptable tolerances
for emittance growth due to both injection jitter and structure misalignments,
for both the nominal (2.8~ns) and alternate (1.4~ns) bunch spacings.
For the L-band structure (a structure with $2\pi/3$ phase advance)
we take a uniform distribution in synchronous
dipole mode frequencies,
with central frequency $\bar{f}=2.05$~GHz and width
$\Delta_{\delta f}=3\%$. 
For the S-band case 
our optimized structure ( a $3\pi/4$ structure) has a trapezoidal 
dipole frequency distribution with
$\bar{f}=3.92$~GHz, $\Delta_{\delta f}=5.8\%$, and tilt parameter
$\alpha=-.2$.
The central frequency and phase advance were
chosen to put bunches early in the train on
the zero crossing of the wake and, at the same time, keep the
gradient optimized.
We have shown that for random manufacturing errors with rms 5~$\mu$m,
(equivalent to $10^{-4}$ error in synchronous frequency),
the injection jitter tolerances are still acceptable.
We have also shown that the structure alignment tolerances are loose,
and that the cell-to-cell misalignment tolerance is $\gtrsim 40$~$\mu$m.
Note that in this report we have considered only the effects of modes in the
first dipole passband. 

\end{quote}

}
\vfill
 
\title{Dipole Mode Detuning in the\break Injector Linacs of the NLC}
\author{Karl L.F. Bane and Zenghai Li}
\maketitle

\section{Introduction}

A major consideration in the design of the accelerator structures in the
 injector linacs of the JLC/NLC\cite{zdr}\cite{NLC}
is to keep the wakefield effects within tolerances for both the nominal
 (2.8~ns)
and the alternate (1.4~ns) bunch spacings.
One important wakefield effect 
in the injector linacs is likely to be multi-bunch beam break-up (BBU).
With this effect a jitter in the injection conditions of a
bunch train, due to the dipole modes of the
accelerator structures, is amplified in the linac.
By the end of the linac
bunches in the train
are driven to large amplitudes and/or the projected emittance
of the train becomes large,
both effects which can hurt machine performance.
Another important multi-bunch wakefield effect 
that needs to be considered is static emittance
growth caused by structure misalignments.

To minimize the multi-bunch wakefield effects
 in the injector linacs we need to minimize the sum wake in
the accelerator structures.
The dipole wake amplitude of the structures---and
therefore also the sum wake amplitude---scales
 as frequency to the -3 power. Therefore, compared to the
main (X--band) linac, the injector linac wakes tend to be smaller by a factor 
$1/64$ and $1/512$, respectively, for 
 the S-- and L--band linacs.
We shall see, however, that---in the S-band case---this
 reduction, by itself, is not sufficient.
Two ways of reducing the sum wake further are
to detune the first pass-band dipole modes
 and to damp them.
Detuning can be achieved by gradually varying the dimensions of the cells in a structure.
Weak damping can be achieved by letting the fields couple to manifolds running
parallel to the structure (as is done in the main JLC/NLC linac\cite{RJones});
 stronger damping by, for 
example,
introducing lossy material in the cells of the structure.

In the injector linacs the dipole mode frequencies are much lower than in
the main linac, and 
the number of dipole mode oscillations between bunches $n_d$ is much smaller
(see Table~\ref{tascale}).
Therefore, significantly reducing the wake envelope at 
one bunch spacing 
behind the driving bunch by detuning alone becomes more difficult.
In addition, for a given $Q$, the effective damping is 4 (or 8) times
less effective than for X-band.

\begin{table}[htb]
\caption{Scaling of the frequency and the wake amplitude for C, 
S, and L bands as compared to
X band. Also given are the number of dipole mode oscillations between bunches $n_d$,
and the damping needed to reduce the wake amplitude by $1/e$
at the position of the second bunch, $Q_d$, for the nominal
(2.8~ns) and the alternate (1.4~ns) bunch spacings.
}
\label{tascale}
\vskip6mm
\centering
\begin{tabular}{||c|c|c||r|r||r|r||} \hline \hline
\multicolumn{3}{||c||}{Scaling}   &  \multicolumn{2}{c||}{$\Delta t=2.8$~ns} & \multicolumn{2}{c||}{$\Delta t=1.4$~ns}    \\ \hline\hline
Band & Freq. & Wake    &  $n_d$ & $Q_d$ & $n_d$  &   $Q_d$   \\ \hline\hline
X &     1  &  1      &   42.0 & 132 & 21.0 & 66   \\ \hline
C &     1/2  &  1/8      &   21.0 & 66 & 10.5 & 33   \\ \hline
S &     1/4  &  1/64      &   10.5 & 33 & 5.3 & 16   \\ \hline
L &     1/8  &  1/512      &   5.3 & 16 & 2.6 & 8   \\ \hline\hline
\end{tabular}
\end {table}

In this report our goal is to design the accelerator structures for the
injector linacs using simple detuning alone, {\it i.e.} including no damping,
to take care of the long-range wakefields.
We focus mostly on the S-band injector linacs.
We begin by discussing analytical approaches to estimating the
effects of BBU and structure misalignments.
We then discuss wakefield compensation using detuning.
We optimize structure dimensions for structures with $2\pi/3$ and
$3\pi/4$ per cell phase advance, and show that the latter is preferable.
And finally we obtain tolerances to wakefield effects
for all the injector linacs using both analytical formulas and
numerical tracking. Note that in this report we are only concerned with the
effects of modes in the first dipole passband, which have kick factors
much larger than those in the higher passbands. The effects of the higher
passband modes, however, will need to be addressed in the future.

\section{Emittance Growth}

\subsection{Beam Break-up (BBU)}

In the case of {\it single-bunch} beam break-up 
in a linac the amplification
of injection jitter can be characterized by a strength parameter
dependent on the longitudinal position within the bunch.
When the strength parameter is sufficiently small
the growth in amplitude at the end of the linac is given by the
first power of this parameter\cite{chao}.
For the {\it multi-bunch} case we can derive an analogous
strength parameter, one dependent on bunch number $m$.
When this strength parameter is sufficiently small
we expect that
again the growth in amplitude at the end of the linac is given by the
first power of the parameter. 
(But even when the strength parameter is not sufficiently small
it can be a useful parameter for characterizing the strength of BBU.)
For the multi-bunch case the strength parameter becomes
(see Appendix~A)
\begin{equation}
\Upsilon_m= {e^2NLS_{m}\bar\beta_0\over 2E_0} 
g(E_f/E_0,\zeta)\quad\quad\quad[m=1,\ldots,M]\
,\label{eqeta}
\end{equation}
with $N$ the single bunch population, $L$ the
machine length, 
$\bar\beta_0$ the initial value of the beta function averaged over a lattice cell,
$E_0$ the initial energy, $E_f$ the final energy, and $M$ the number
of bunches in a train.
The sum wake $S_m$ is given by
\begin{equation}
S_m= \sum_{i=1}^{m-1} W[(m-i)\Delta t]\quad\quad\quad[m=1,\ldots,M]\ ,
\end{equation}
with $W$ the transverse wakefield
and $\Delta t$ the time interval between bunches in a train.
The wakefield, in turn, is given by a sum over the dipole modes
in the accelerator structures:
\begin{equation}
W(t)= \sum^{N_m}_n 2 k_{n}\sin({2\pi f_{n}t/ c})
\exp(-\pi{f_{n}}t/Q_n)\quad\quad,
\label{eqwake}
\end{equation}
with $N_m$ the number of modes, $f_n$, $k_n$, and $Q_n$ are, respectively,
the frequency, the kick factor, and the quality factor of the $n^{\rm th}$
mode.
The function $g(x)$ in Eq.~\ref{eqeta} is one depending on the
energy gradient and focusing profile in the linac.
For acceleration 
assuming the beta function varies as $\bar\beta\sim E^\zeta$,
\begin{equation}
g(x,\zeta)= {1\over\zeta}\left({x^\zeta-1\over x-1}\right)
\quad\quad\quad[{\bar\beta\sim E^\zeta}].
\end{equation}

If $\Upsilon_m$, for all $m$, is not large, the linear approximation
applies, and this parameter
directly gives
 the (normalized) growth in amplitude of bunch $m$.
If $\Upsilon_m$ is not large
the projected normalized emittance growth of the bunch train
becomes (assuming, for simplicity,  that, in phase space, the beam ellipse
is initially upright):
\begin{equation}
\delta\epsilon\approx 
\left[{1+\left({y_0\Upsilon_{rms0}\over 
\sigma_{y0}}\right)^2}\right]^{1/2}-1\quad\quad\quad [\Upsilon_m\ {\rm small}],\label{eqemit}
\end{equation}
with $y_0$ the initial bunch offset, $\Upsilon_{rms0}$
the rms of the strength parameter
(the square root of the second moment:
 the average is not subtracted), and $\sigma_{y0}$ the initial
beam size.
Note that
the quantity $S_m/M$ in the multi-bunch case
takes the place
of the bunch wake (the convolution of the wake with the
bunch distribution)
in the single bunch instability problem.
As jitter tolerance parameter, $r_t$, we can take that ratio $y_0/\sigma_{y0}$
that yields a tolerable emittance growth, $\delta_{\epsilon t}$.

\subsection{Misalignments}

If the structures in the linac are (statically) misaligned with respect
to a straight line, the beam at the end of the linac will have an 
increased projected emittance.
If we have an ensemble of misaligned linacs then,
to first order, the 
distribution in emittance growth at the end of these linacs
is given by an exponential distribution
$\exp[-\delta\epsilon/\langle\delta\epsilon\rangle]/\langle\delta\epsilon\rangle$, 
with\cite{static}
\footnote{This equation is a slightly generalized
form of an equation given in Ref.~\cite{static}.}
\begin{equation}
\sqrt{\langle\delta\epsilon\rangle}= {e^2NL_a(x_a)_{rms}{S}_{rms}\over E_0}
\sqrt{{N_a\beta_0\over2}}\,h(E_f/E_0,\zeta) 
\quad,
\label{eqmisa}
\end{equation}
with $L_a$ the structure length, 
$(x_a)_{rms}$ the rms of the structure misalignments,
${S}_{rms}$ is the rms of the 
sum wake {\it with respect to the average},
$N_a$ the number of structures;
the function $h$ is given by
(again assuming $\bar\beta\sim E^\zeta$):
\begin{equation}
h(x,\zeta)= \sqrt{{1\over\zeta x}\left({x^\zeta-1\over x-1}\right)}
\quad\quad\quad[{\bar\beta\sim E^\zeta}].
\end{equation}
Eq.~\ref{eqmisa} is valid assuming the so-called betratron term
in the equation of motion is small compared to the misalignment term.

We can define a misalignment tolerance by 
\begin{equation}
x_{at}=(x_a)_{rms}\sqrt{{\delta\epsilon_t\over\langle\delta\epsilon\rangle}}\quad,
\label{eqxat}
\end{equation}
 with $\delta\epsilon_t$ the tolerance in emittance growth.
What is the meaning of $x_{at}$?
For an ensemble of machines, each with a different collection of
random misalignment errors but with the same rms $x_{at}$, then the 
distribution of final emittances
will be given by the exponential function 
 with expectation value
 $\delta\epsilon_t$. Note that if we,
for example,
want to have 95\% confidence to achieve this emittance growth, we
need to align the machine to a tolerance level of
 $x_{at}/\sqrt{-\ln .05}\approx.58x_{at}$.

Besides the tolerance to structure misalignments, we are also interested
in the tolerance to cell-to-cell misalignments
due to fabrication errors. A structure is built
as a collection of cups, one for each cell, that are brazed together,
and there will be some error, small compared
to the cell dimensions, in the straightness of each structure.
To generate
a wake (for a beam on-axis) in a structure 
with cell-to-cell misalignments
we use a perturbation approach that assumes 
that, to first order,
the mode frequencies remain unchanged (from those in the straight structure),
and only new kick factors are needed\cite{perturb} 
(The method is described in more detail in Appendix~B).
Note that for particle tracking through structures with internal
misalignments, contributions from both this (orbit independent) wake force
and the normal (orbit dependent) wake force need to be included.

\noindent\rule{12.6cm}{.5mm}
\vspace{3mm}

Machine properties for the injector linacs used in this report are given in 
Table~\ref{taone}\cite{NLC}.
The rf frequencies of all linacs are sub--harmonics of the main linac
frequency, 11.424~GHz.
The prelinac, $e^+$ drive linac, $e^-$ booster linac all operate at
S--band (2.856~GHz), and the $e^+$ booster linac at L--band (1.428~GHz).
Note that $\bar\beta_{y0}$ and $\zeta$ are only a rough fitting of the
real machine $\beta$--function to the dependence $\bar\beta\sim E^\zeta$.
In Table~\ref{tatwo} beam properties for the injector linacs, for the nominal
bunch train configuration (95 bunches spaced at $\Delta t=2.8$~ns),
are given.
For the alternate configuration (190 bunches spaced at $\Delta t=1.4$~ns)
$N$ is reduced by $1/\sqrt{2}$.

\begin{table}[htb]
\caption{Machine properties of the injector linacs. Given are the initial energy $E_0$,
the final energy $E_f$, the length $L$, the initial average beta function in $y$,
and the approximate scaling parameter $\zeta$, of $\beta$ with energy ($\beta\sim E^\zeta$).}
\vskip6mm
\centering
\label{taone}
\begin{tabular}{||l||c|c|c|c|c|c||} \hline \hline
Name   &  Band &$E_0$[GeV]  & $E_f$[GeV] & $L$[m]  &  $\bar\beta_{y0}$[m] &  $\zeta$   \\ \hline\hline
Prelinac & S &    1.98   &  10.0     & 558    &   8.6               &    1/2      \\ \hline
$e^+$ Drive & S &  .08   &  6.00     & 508    &   2.4               &    1/2      \\ \hline
$e^-$ Booster & S & .08   &  2.00     & 163    &   3.4               &    1/4       \\ \hline
$e^+$ Booster & L &.25   &  2.00     & 184    &   1.5               &     1       \\ \hline\hline
\end{tabular}
\end {table}

\begin{table}[htb]
\caption{Beam properties in the injector linacs under the nominal
bunch train configuration (95 bunches spaced at $\Delta t=2.8$~ns).
 Given are the bunch population $N$,
the rms bunch length $\sigma_z$, the initial energy spread $\sigma_{\delta 0}$, and
the nominal normalized emittance in $y$, $\epsilon_{yn}$.
Note that under the alternate bunch train configuration (190 bunches spaced
at $\Delta t=1.4$~ns) $N$ is reduced by $1/\sqrt{2}$.
}
\vskip6mm
\centering
\label{tatwo}
\begin{tabular}{||l||c|c|c|c||} \hline \hline
Name    &  $N[10^{10}]$ & $\sigma_z$[mm] & $\sigma_{\delta 0}$[\%]  &   $\epsilon_{yn}$[m]   \\ \hline\hline
Prelinac &     1.20   &  0.5      &   1.                &  $3\times10^{-8}$    \\ \hline
$e^+$ Drive &  1.45   &  2.5      &   1.                &  $1\times10^{-4}$    \\ \hline
$e^-$ Booster &1.45   &  2.5      &   1.                &  $1\times10^{-4}$    \\ \hline
$e^+$ Booster &1.60   &  9.0      &   3.5               &  $6\times10^{-2}$    \\ \hline\hline
\end{tabular}
\end {table}

\section{Wakefield Compensation}

For effective detuning,
one generally requires that the wake amplitude drop quickly, in the
time interval between the first two bunches, and then remain low until the tail
of the bunch train has passed.
In the main (X-band) linac of the NLC, Gaussian detuning is used to 
generate a fast Gaussian fall-off in the wakefield; in particular,
at the position of the second bunch the wake is reduced by roughly 2 orders of magnitude
from its initial value.
The short time behavior of the wake can be analyzed
by the so-called ``uncoupled'' model. According to this model
(see, for example, Ref.~\cite{Gluck})
\begin{equation}
W(t)\approx \sum^{N_c}_n 2 k_{sn}\sin({2\pi f_{sn}t/ c})
\quad\quad\quad[t\ {\rm small}],
\label{eqwakes}
\end{equation}
where $N_c$ is the number of cells in the structure,
and $f_{sn}$ and $k_{sn}$ are, respectively, the frequency
and kick factor at the synchronous point, for a periodic
structure with dimensions of cell $n$.
Therefore, one can predict the short time behavior of the wake without
solving for the modes of the system.
(In the following we will omit the unwieldly subscript $s$; whether the
synchronous or mode parameters are meant will be evident from context.)

For Gaussian detuning the initial fall-off of the wake is given by
\begin{equation}
W(t)\approx 2{\bar k}\sin(2\pi{\bar f}t)
\exp\left(-2\left[\pi{\bar f} t\sigma_{\delta f}\right]^2\right)
\quad\quad\quad[t\ {\rm small}],
\end{equation}
with ${\bar k}$ the average kick factor, ${\bar f}$ the average 
(first band) synchronous, dipole mode frequency,
 and $\sigma_{\delta f}$ the sigma parameter in the Gaussian distribution.
Suppose we want a relative amplitude reduction to 0.05 at the position of the second bunch.
 Considering the alternate (1.4~ns) bunch
spacing, and taking $\bar{f}=4.012$~GHz (S-band), we find that the required
$\sigma_{\delta f}=6.5\%$. To achieve a smooth Gaussian drop--off of the wake requires that
we take at least $\sim 3\sigma_{\delta f}=20\%$ as the full--width of our frequency
distribution, a number which is clearly too large.

If we limit the total frequency spread to 
an acceptable $\Delta_{\delta f}=10\%$ and keep the parameter 
$\sigma_{\delta f}$ fixed, our Gaussian
distribution becomes similar to a uniform distribution. For the case of a uniform
distribution with full width $\Delta_{\delta f}$
the wake becomes
\begin{equation}
W\approx {2\bar{k}\over N_c}
\sin(2\pi{\bar f}t)\,{{\rm sin}(\pi{\bar f}t\Delta_{\delta f})\over
  {\rm sin}(\pi{\bar f}t\Delta_{\delta f}/N_c)}\quad\quad\quad
[(\pi\bar{f}t/Q)\,{\rm small}].\label{equni}  
\end{equation}

Again considering S-band with the alternate (1.4~ns) bunch spacing, and taking 
$\bar f=4.012$~GHz, we obtain an amplitude reduction to 
0.37 at the position of the second bunch, which is still too large.
If we want to substantially reduce the wake further we need to shift the 
average dipole frequency ${\bar f}$, so that
the term $\sin(2\pi{\bar f}t)$ 
in Eq.~\ref{equni} becomes small, and the wake at the second
bunch is near a zero crossing. That is,
\begin{equation}
\bar{f}\Delta t={n\over2}\quad\quad\quad[n\,{\rm an\, integer}],
\label{eqcond}
\end{equation}
with $\Delta t$ the bunch spacing.
With $n$ an even integer the bunch train will be near the integer resonance,
otherwise it will be near the half-integer resonance.
With our parameters $\bar{f}\Delta t=5.62$, and 
Condition \ref{eqcond} is 
achieved by changing ${\bar f}$ by $-2\%$ (or by a much larger
amount in the positive direction).
However, $\bar f=4.012$~GHz is the average dipole mode
frequency for the somewhat optimized structure,
and a change of $-2\%$ results in a net loss
of 7\% in accelerating gradient,
 and, presumably, a 7\% increase in the required
lengths of the S--band injector linacs.
One final possibility for reducing the wake at one bunch spacing is to introduce
heavy damping. But for this case, just to reduce the wake at one bunch spacing by $1/e$,
a quality factor of 16 would be needed (see Table~\ref{tascale}), and
 such a quality factor is not easy to achieve without 
a significant loss in
fundamental mode shunt impedance.

\begin{figure}[htb]
\centering
\epsfig{file=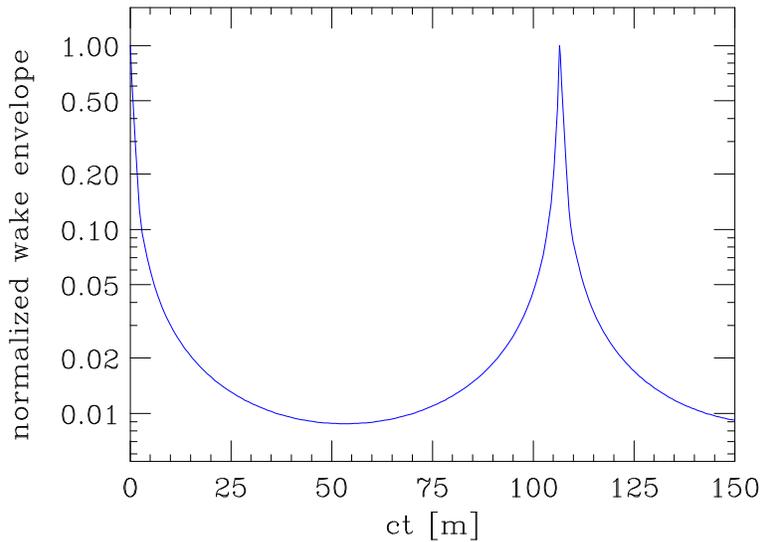, width=10cm}
\caption{The wake envelope (normalized to a maximum of 1)
for a uniform frequency distribution.
Shown is Eq.~\ref{equni} with all oscillations removed.
The average frequency $\bar{f}=4.012$~GHz, the bandwidth $\Delta_{\delta f}=8\%$,
and the number of modes $N_c=114$. Note that in the injector linacs 
the bunch train extends to $ct=80$~m.
}
\label{fisinxox}
\end{figure}

The wakefield for a uniform distribution, as given by Eq.~\ref{equni},
not only gives the initial drop-off of the wake, but also the longer
term behavior. (However, here the mode parameters, not the synchronous
parameters, are needed. Therefore, to see whether such a mode distribution
can actually be achieved the circuit equations need to be solved.)
We see that for a uniform
distribution the wakefield resurges to a maximum again,
at $t=N_c/({\bar{f}\Delta_{\delta f}})$. Therefore, $\Delta_{\delta f}$ must be
sufficiently small to avoid this resurgence occurring 
 before the end of the bunch train; {\it i.e.} 
it must be significantly less than
$N_c/(M{\bar f}\Delta t)$ (which is about 10\% in our case).
The envelope of
Eq.~\ref{equni} for $\Delta_{\delta f}=8\%$, ${\bar f}=4.012$~GHz,
and $N_c=114$
is shown in Fig.~\ref{fisinxox}. 

Another possibility for pushing the resurgence in the wake 
to larger $t$ is to use two
structure types, which can effectively double the number of modes available for
detuning.
This idea has been studied; 
it has been rejected in that it requires extremely tight alignment
tolerances between pairs of such structures.

\subsection{$2\pi/3$ Phase Advance Per Cell}

Except for the region of the initial drop-off, we need to solve for
the eigenmodes of the system to know the behavior of the wake
or the sum wake for a detuned structure.
To numerically obtain these modes
we use a computer program that solves the double-band circuit model described
in Ref.~\cite{Gluck}. 
We consider structures of the disk--loaded type, with constant period and
with rounded irises of fixed thickness.
The iris radii and cavity
radii are adjusted to give the correct fundamental mode frequency and the desired
dipole mode spectrum.
Therefore, the dimensions of a particular cell $m$ can be specified by one free
parameter, which we take to be the synchronous frequency of the first
dipole mode pass band, $f_{sm}$ (more precisely, the synchronous frequency of
the periodic structure with cell dimensions of cell $m$). 
The computer program generates $2N_c$
coupled mode frequencies $f_{n}$ and kick factors $k_{n}$,
with $N_c$ the number of cells in a structure. 
It assumes the modes are trapped at the ends of the structure.
Only the modes of the first band (approximately the first $N_c$ modes)
are found accurately by the two-band model.
And since, in addition, the strengths of the first band modes
are much larger
than those of the second band (in the S-band case the synchronous mode
kick factors are larger by a factor $\sim35$),
we will use only the
first band modes to obtain the wake,
and then the sum wake.

For our S-band structures, 
we will consider a uniform frequency distribution, with a central frequency
$\bar{f}$ chosen so that at one bunch spacing, for the alternate configuration
($\Delta t=1.4$~ns), the wake is very close to a zero crossing. 
The strength of interaction with the modes, given by the kick factors $k$,
will be stronger near the downstream
end of the structure, where the iris radii become
smaller. To counteract this asymmetry we will allow
the top of the frequency distribution to be slanted at an angle, and
therefore, our distribution becomes trapezoidal in shape.
We parameterize this slant by
\begin{equation}
\alpha= {\lambda_f(f_{hi})-\lambda_f(f_{lo})\over\lambda_f(f_{hi})+\lambda_f(f_{lo})}\quad\quad,
\end{equation}
where $\lambda_f$ is the synchronous 
frequency distribution, and $f_{lo}$ and $f_{hi}$ represent,
respectively, the lowest and highest frequencies in the distribution.
Note that $-1\leq\alpha\leq1$.
With $\bar f$, $\alpha$, and 
the relative 
width of the distribution $\Delta_{\delta f}$, we have 3 parameters
that we will vary to reduce the wakefield effects---specifically by minimizing
on the sum wake---for both bunch train configurations.

\begin{figure}[htbp]
\centering
\epsfig{file=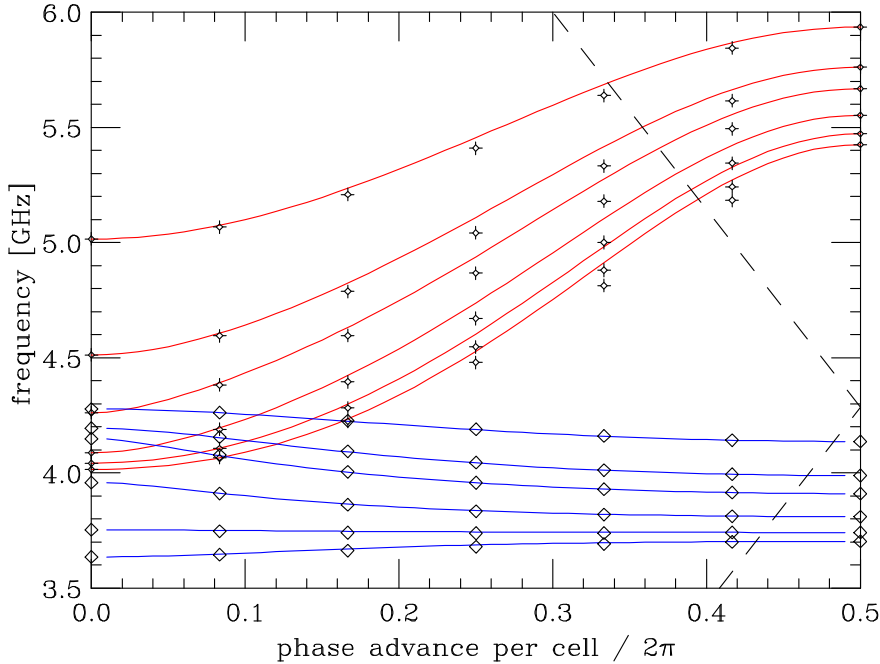, width=11.6cm}
\caption{
The dispersion curves of the first two dipole bands
of representative constant impedance 
 S-band structures. The phase advance per cell is $2\pi/3$.
Results are given for iris radii of $a=1.30$, 1.51, 1.63, 1.80, 1.92, and 2.00~cm.
The plotting symbols give OMEGA2 results, the curves those of the
circuit model. The dashed line is the speed of light curve.
}
\label{fiomegbet}
\end{figure}

Each S-band structure 
operates at a fundamental mode frequency of 2.856~GHz, and
consists of 114 cells with a cell period of 3.5~cm (where the phase advance per cell
$\phi=2\pi/3$), an iris thickness of 0.584~cm, and cavity radius $\sim4.2$~cm.
The $Q$ of the modes due to wall losses (copper) $\sim14,500$.
Given our implementation of the SLED-I pulse
compression system\cite{zenghai}, to optimize the rf efficiency
the average (synchronous) dipole mode frequency
needs to be 4.012~GHz. 
Fig.~\ref{fiomegbet} shows the dispersion curves of the first two dipole
bands for representative constant impedance, S-band
 structures, with $a$ varying
from 1.30~cm to 2.00~cm. The results of a 
finite element, Maxwell Equations solving program,
OMEGA2\cite{OMEGA2}, are given by the plotting
symbols. The end points of the curves are used to fix the parameters in the
circuit program. The two-band 
circuit results for these constant impedance structures
are indicated by the curves in the figure. We note good
agreement in the first band results
and not so good agreement in those of the second band.
 For a detuned structure, to obtain
the local circuit parameters, we interpolate from these representative 
dispersion curves.

We have 3 parameters to vary in our input (uncoupled) frequency distribution:
the (relative) shift in average frequency from the nominal 4.012~GHz,
$\delta\bar{f}$, the (relative) width of the distribution $\Delta_{\delta f}$,
and the flat-top tilt parameter $\alpha$.
Varying these parameters
we calculate $S_{rms0}$, $S_{rms}$, and 
the peak value of $|S|$, $|\hat S|$, for the coupled results,
for both bunch train configurations.
These parameters serve as indicators, respectively, of
emittance growth due to BBU (injection jitter),
 emittance growth due to misalignments, and the
maximum beam excursion due to BBU.
From our numerical simulations  
we find that a fairly optimized case consists of
$\delta\bar{f}=-2.4\%$, $\Delta_{\delta f}=7.5\%$, and $\alpha=-0.20$.

\begin{figure}[htbp]
\centering
\epsfig{file=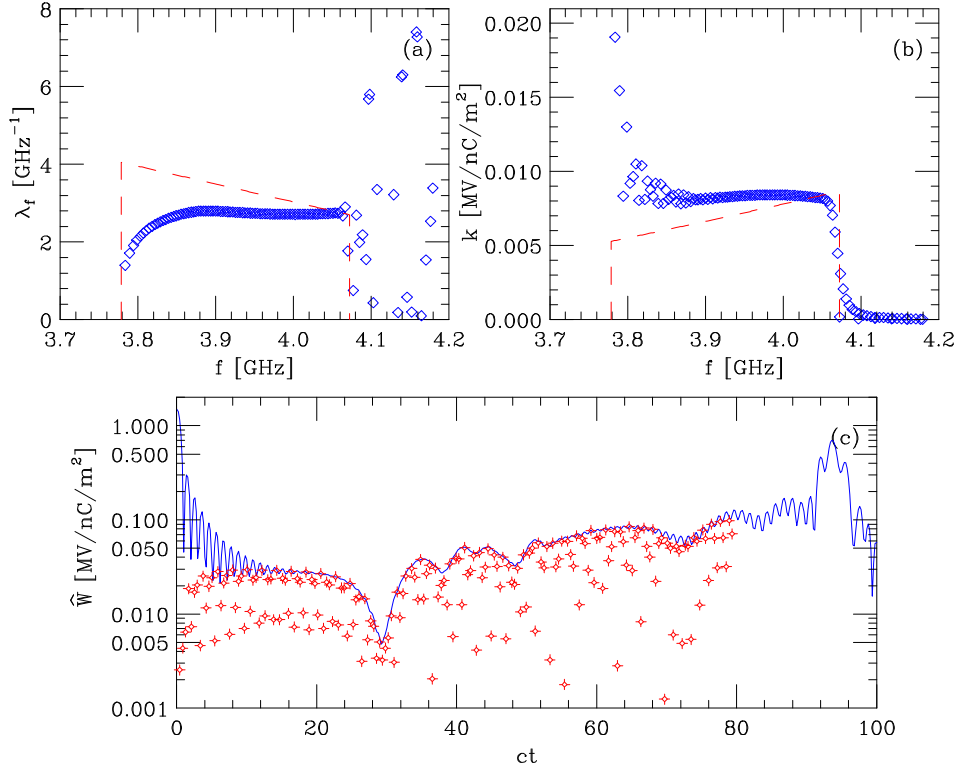, width=12.6cm}
\caption{
The optimized $2\pi/3$ structure:
 $\delta\bar{f}=-2.4\%$, $\Delta_{\delta f}=7.5\%$, and $\alpha=-0.20$.
Given are
the frequency distribution $\lambda_f$~(a), the mode kick factors $k$~(b), and
the envelope of the wake $\hat W$~(c). The dashes in (a) and (b) give the uncoupled
results; the plotting symbols in (c) give $|W|$ at the bunch positions for
the alternate (1.4~ns) bunch train configuration. Note that we display only 
the modes of the first dipole band, and the wake due to these modes.
}
\label{fisband_opt}
\end{figure}

\begin{figure}[htbp]
\centering
\epsfig{file=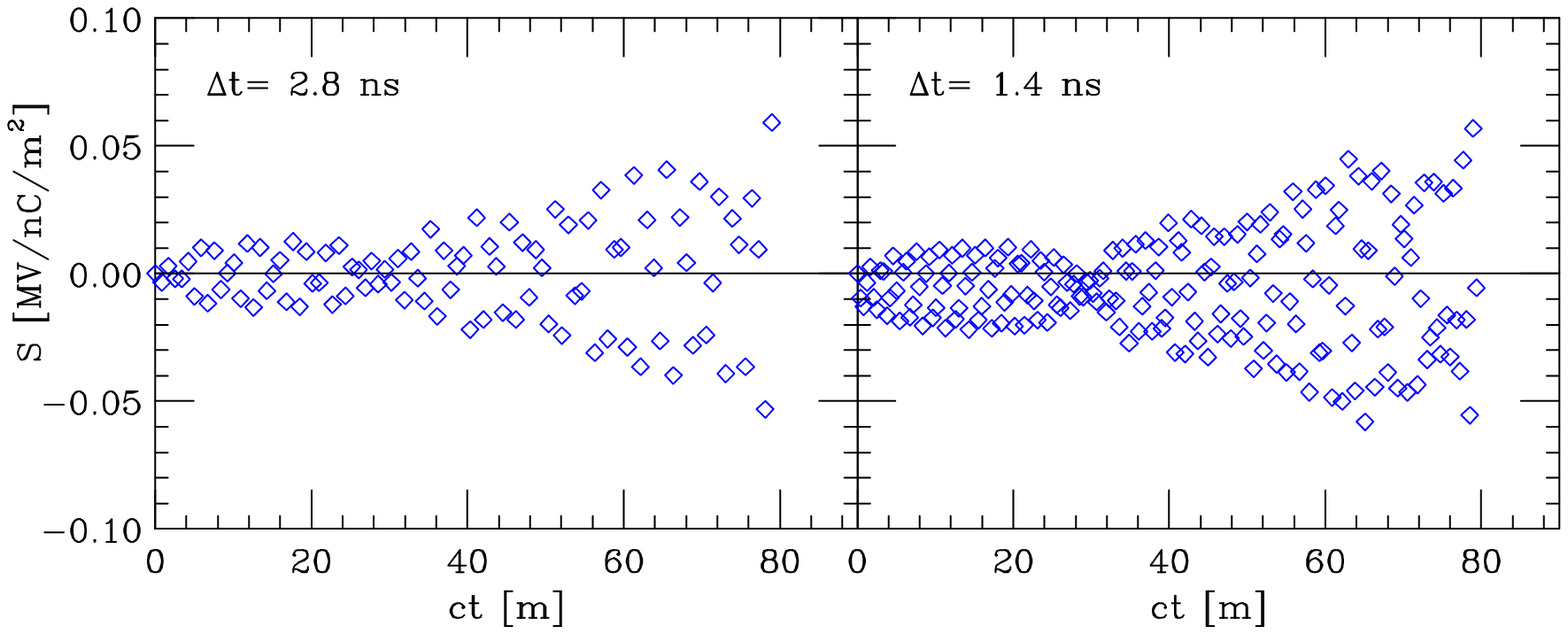, width=12.6cm}
\caption{
The sum wake for the optimized $2\pi/3$ structure,
for both the nominal and
alternate bunch train configurations.
In both cases $S_{rms0}=S_{rms}=.02$~MV/nC/m$^2$.}
\label{fisbands_opt}
\end{figure}

In Fig.~\ref{fisband_opt} we display, for the optimized case, 
the frequency distribution~(a), the kick factors~(b), and the envelope of the wake~(c).
The dashed curves in (a) and (b) give the uncoupled (input) values.
The plotting symbols in (c) give $|W|$ at the bunch positions for
the alternate (1.4~ns) bunch train configuration. 
From (b) we note that there are a few modes, trapped near the beginning of the
structure, which have kick factors significantly larger than the rest.
This is a consequence 
of the fact that, for all cells of this structure, the dispersion
curves are backward waves. 
From (c) we see that, due to these few strong modes,
the wake envelope does not nearly 
reach the low, flat bottom that it does for the idealized,
uniform frequency distribution (see Fig.~\ref{fisinxox}). 
We note, however, that the short-range drop-off is similar to
 the idealized form 
(see Fig.~\ref{fisinxox}), for about 20~m.
In addition we note that,
by setting the second bunch near the zero crossing, many following bunches
also have wakes with amplitudes significantly
below the wake envelope.
Finally, in Fig.~\ref{fisbands_opt} we present the sum wake for both bunch
train configurations.
For this case, for both bunch train configurations,
$S_{rms0}=S_{rms}=.02$~MV/nC/m$^2$.
Note that if we set $\delta{\bar f}$ back to 0, then, for the 1.4~ns bunch 
spacing option, $S_{rms0}$ becomes a factor of 20 larger. 

\subsection{$3\pi/4$ Phase Advance Per Cell}

\begin{figure}[htbp]
\centering
\epsfig{file=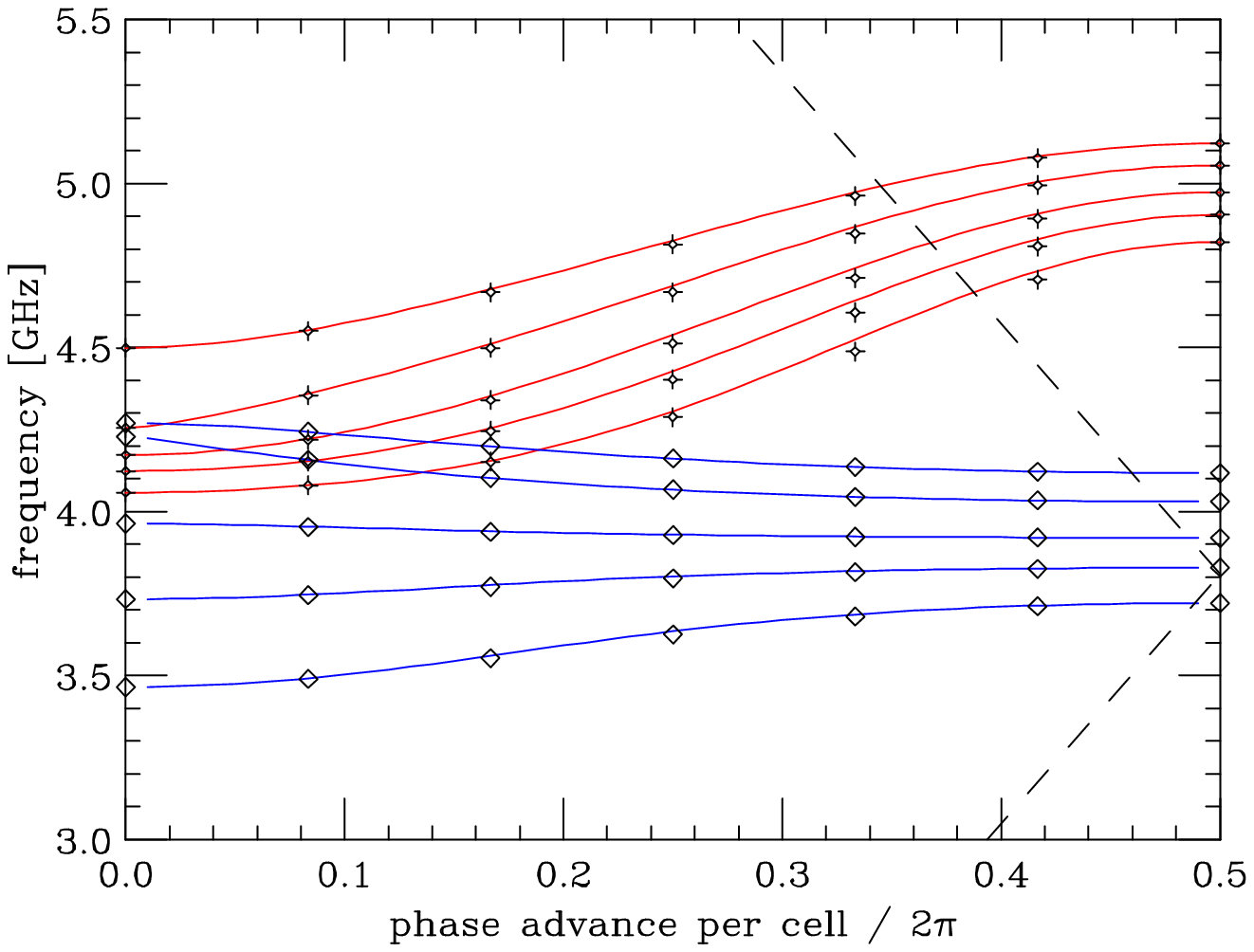, width=11.6cm}
\caption{
The dispersion curves of the first two dipole bands
of representative cells in a
$3\pi/4$ structure. 
Results are given for iris radii of $a=1.33$, 1.48, 1.63, 1.80, and 1.98~cm.
The plotting symbols give OMEGA2 results, the curves those of the
circuit model. The dashed line is the speed of light line.
}
\label{fiomegbet_3pi4}
\end{figure}

If we would like to regain some of the 7\% in accelerating gradient
that we lost by shifting $\bar f$, we
can move to a structure  where the group velocity 
at the synchronous point is
less than for the $2\pi/3$ structure (for the same $\bar f$).
One solution is to go to a structure with a
$3\pi/4$ synchronous point.
Note that in such a structure
the cell length is 3.94~cm and that there are 102 cells per structure.
Note also that for the same group velocity for the fundamental mode
a higher phase advance implies larger values of iris radius $a$,
which will also improve the short-range wakefield tolerances.
 The dispersion curves
are shown in Fig.~\ref{fiomegbet_3pi4}.
Note that, in this case, our distribution will have
$f_0<f_{\pi}$ for the cell geometries near the beginning
of the structure, $f_0>f_{\pi}$ for the cell geometries near the
end of the structure, while the synchronous phase is near pi phase advance.
Consequently modes touching either end of the structure will only
weakly interact with the beam (see, {\it eg} Ref.~\cite{Gluck}), allowing
us to have a smoother impedance function, and therefore a more uniformly
suppressed wakefield envelope.
This was not the case for $2\pi/3$ structure,
 where the dispersion curves for all cells 
have a negative slope (between 0 and $\pi$ phase advance) 
(see Fig.~\ref{fiomegbet}); it is also not the case for
a $5\pi/6$ structure, where the slopes would all be positive.

\begin{figure}[p]
\centering
\epsfig{file=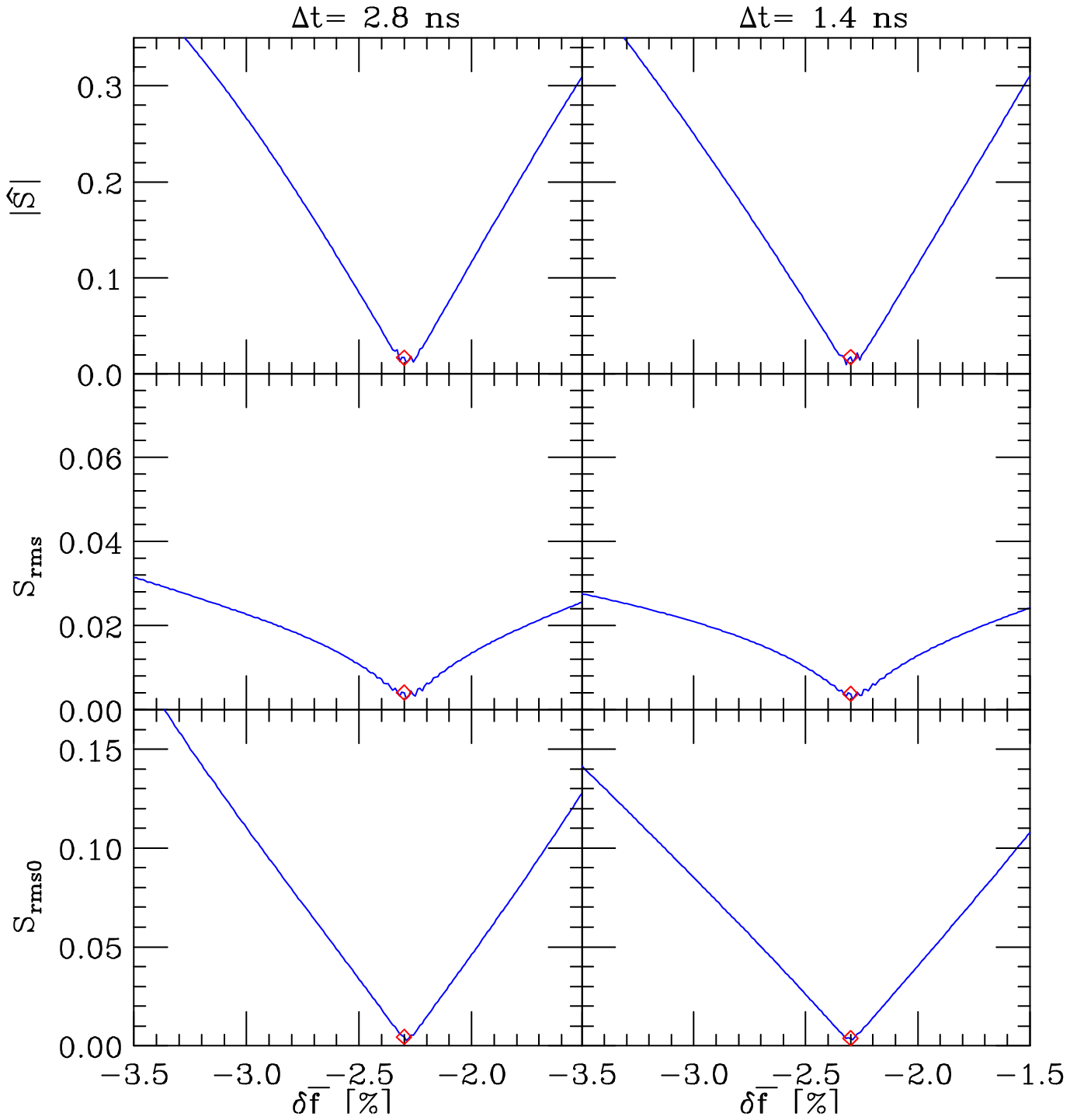, width=12.6cm}
\caption{
For the $3\pi/4$ structure:
dependence of 3 sum wake parameters on
 $\delta\bar{f}$
for the nominal (left frames) and alternate (right frames) bunch
train configurations.
The ordinate units are MV/nC/m$^2$.
The optimum, $\delta\bar{f}=-2.3\%$,
$\Delta_{\delta f}=5.8\%$, and $\alpha=-0.20$, is indicated by the plotting symbol.
}
\label{fivarfbar_3p4}
\end{figure}

\begin{figure}[p]
\centering
\epsfig{file=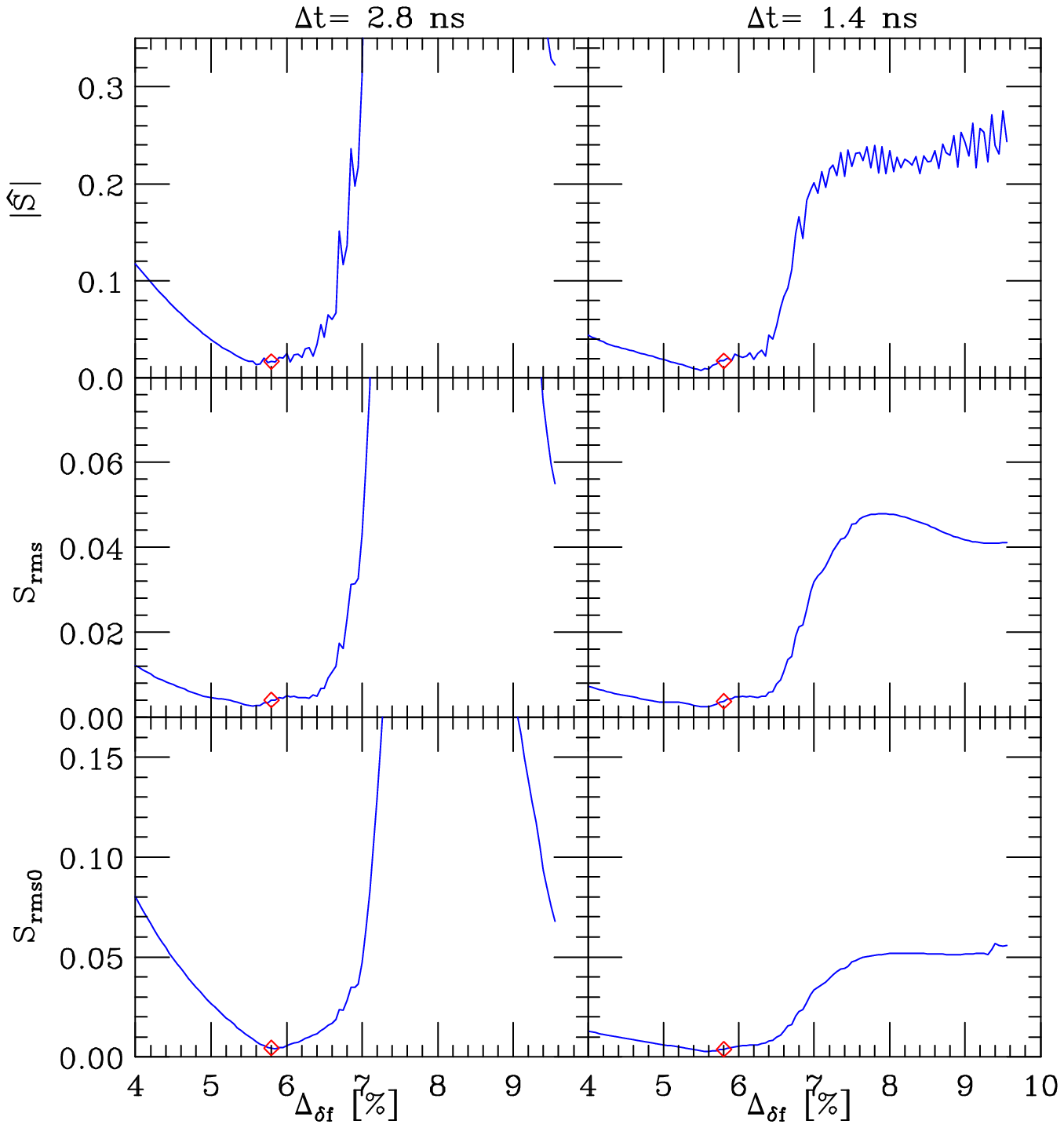, width=12.6cm}
\caption{
For the $3\pi/4$ structure:
dependence of 3 sum wake parameters on
 $\Delta_{\delta f}$, 
for the nominal (left frames) and alternate (right frames) bunch
train configurations.
The ordinate units are MV/nC/m$^2$.
The optimum, $\delta\bar{f}=-2.3\%$,
$\Delta_{\delta f}=5.8\%$, and $\alpha=-0.20$, is indicated by the plotting symbol.
}
\label{fivarftot_3p4}
\end{figure}

\begin{figure}[p]
\centering
\epsfig{file=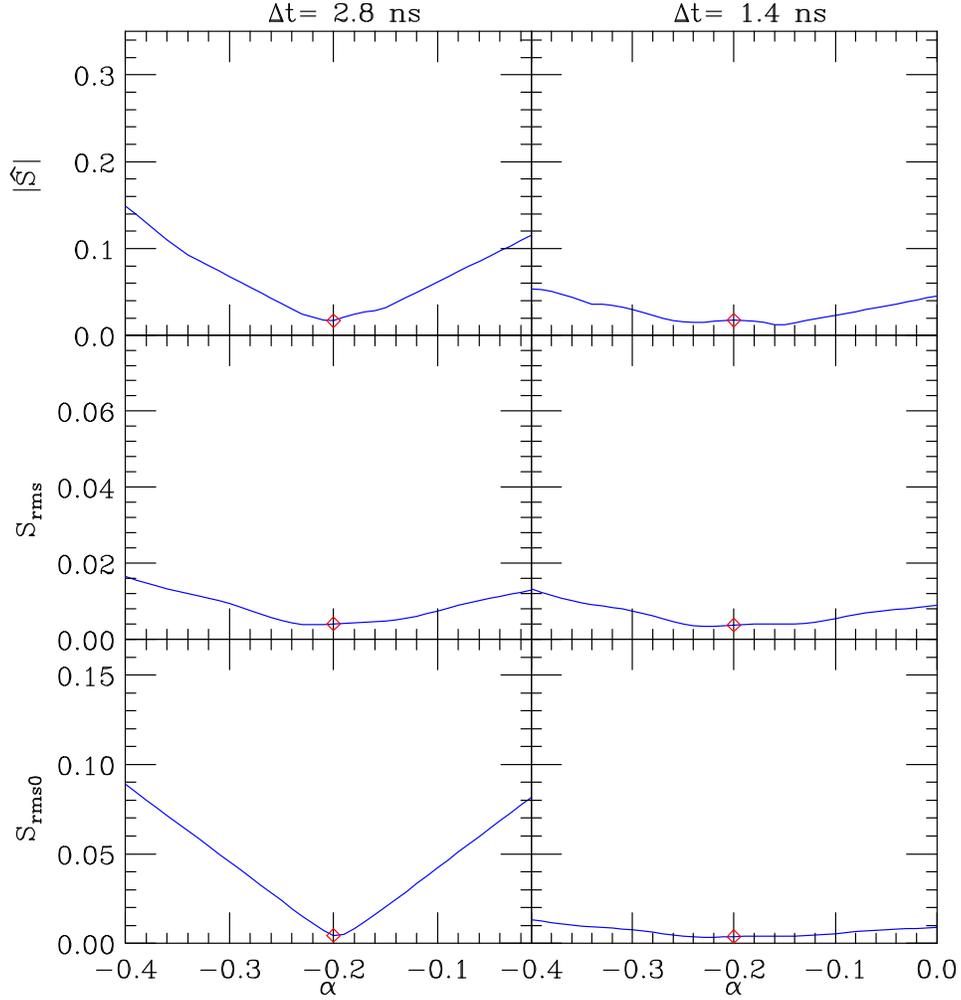, width=12.6cm}
\caption{
For the $3\pi/4$ structure:
dependence of 3 sum wake parameters on
 $\alpha$, 
for the nominal (left frames) and alternate (right frames) bunch
train configurations.
The ordinate units are MV/nC/m$^2$.
The optimum, $\delta\bar{f}=-2.3\%$,
$\Delta_{\delta f}=5.8\%$, and $\alpha=-0.20$, is indicated by the plotting symbol.
}
\label{fivaralpha_3p4}
\end{figure}

\begin{figure}[htbp]
\centering
\epsfig{file=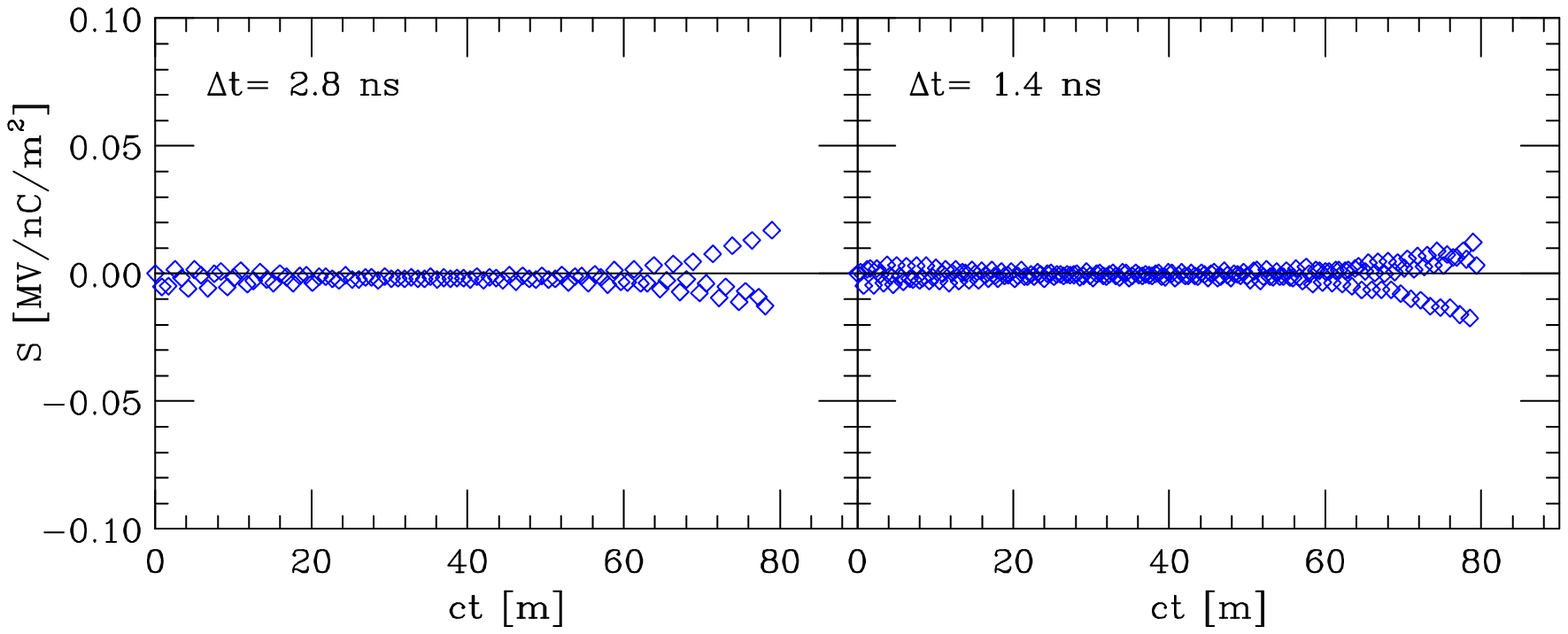, width=12.6cm}
\caption{
The sum wake for the optimized $3\pi/4$ structure,
 given for both the nominal and
alternate bunch train configurations.
For both cases $S_{rms0}=S_{rms}=.004$~MV/nC/m$^2$.}
\label{fisbands_opt_3pi4}
\end{figure}

Again optimizing on the sum wake, we find that for 
a fairly optimized
case  
$\delta\bar{f}=-2.3\%$, $\Delta_{\delta f}=5.8\%$, and $\alpha=-0.20$.
The change of the indicators 
$|\hat S|$, $S_{rms}$ and $S_{rms0}$
as we deviate from this point, for both
bunch train configurations, is shown in 
Figs.~\ref{fivarfbar_3p4}-\ref{fivaralpha_3p4}.
In Fig.~\ref{fivarfbar_3p4} we show the $\bar f$ dependence.
We see that, for both bunch train configurations the results are very sensitive
to $\bar f$.
In Fig.~\ref{fivarftot_3p4} we give the $\Delta_{\delta f}$ dependence.
We can clearly see the effect of the resurgence in the wake
when $\Delta_{\delta f}\gtrsim7\%$. And finally, 
in Fig.~\ref{fisbands_opt_3pi4} we
give the $\alpha$ dependence. We note that the tilt in the distribution
helps primarily in reducing the sensitivity to BBU for the nominal (2.8~ns) bunch
train configuration.

\begin{figure}[htbp]
\centering
\epsfig{file=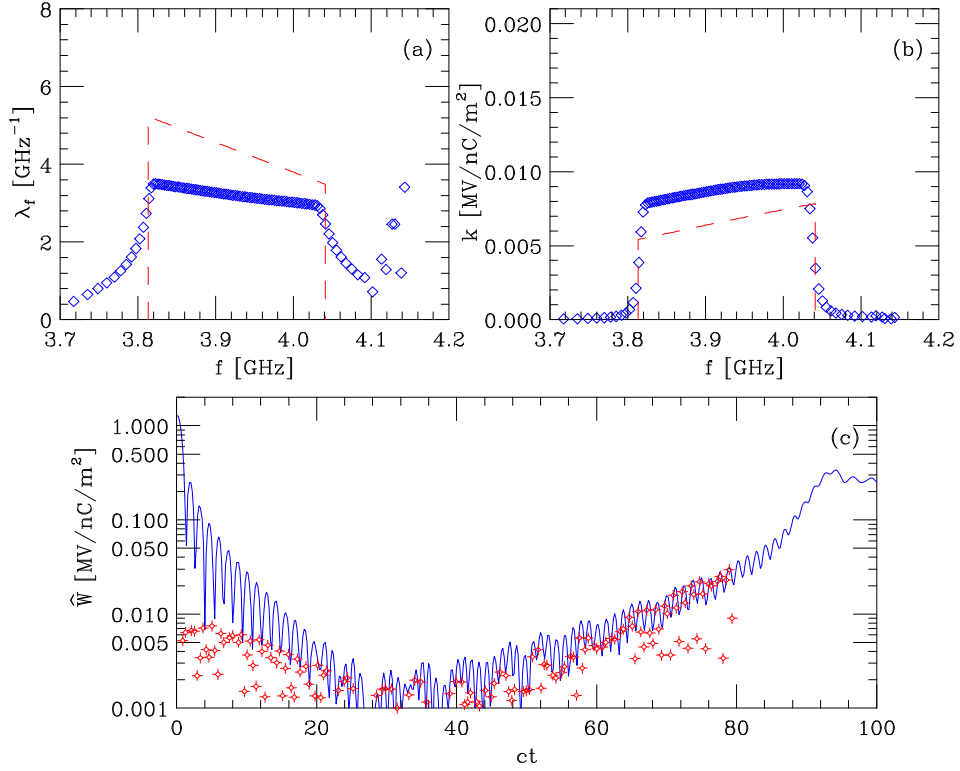, width=12.6cm}
\caption{
For the optimized $3\pi/4$ structure,
 $\delta\bar{f}=-2.3\%$, $\Delta_{\delta f}=5.8\%$, and $\alpha=-0.20$:
the frequency distribution $\lambda_f$~(a), the mode kick factors $k$~(b), and
the envelope of the wake $\hat W$~(c). The dashes in (a) and (b) give the uncoupled
results; the plotting symbols in (c) give $|W|$ at the bunch positions for
the alternate (1.4~ns) bunch train configuration. Note that we display only 
the modes of the first dipole band, and the wake due to these modes.
}
\label{fisband_opt_3p4}
\end{figure}

In Fig.~\ref{fisband_opt_3p4} we display, for the optimized $3\pi/4$ case, 
the frequency distribution~(a), the kick factors~(b), and the envelope of the wake~(c).
From (b) we note that in this case $k(f)$ is a relatively smooth function,
as was expected from our earlier discussion.
From (c) we see that 
the wake envelope reaches a broader, flatter bottom than for the
$2\pi/3$ structure, again as we expected.
Again we note that many of the earlier bunches
have wakes with amplitudes significantly
below the wake envelope.
Finally, in Fig.~\ref{fisband_opt_3p4} we show the sum wake for both bunch
train configurations. The rms of these sum wakes are much smaller than 
for the $2\pi/3$ structure:
$S_{rms0}=S_{rms}=.004$~MV/nC/m$^2$.

\subsection{Frequency Errors}

How sensitive are our results to manufacturing errors? We will begin to
explore this question by looking at the dependence of the sum wake on 
errors in the synchronous frequencies of the cells of the structure.
Note that the synchronous frequency of a cell is not equally sensitive to each
of the cell dimensions. Basically there are 4 dimensions: the iris radius $a$,
the cavity radius $b$, the iris thickness $d$, and the period length $p$.
The synchronous frequency $f_s$ is insensitive to $d$ and $p$, and
for the average S-band cell we find that $\delta f_s=-.85\delta b$
and $\delta f_s=-.15\delta a$. 
Or, a $-1$~micron change in $b$ results in $\delta f_s=2\times10^{-5}$;
a $-1$~micron change in $a$ results in $\delta f_s=1\times10^{-5}$.
As for attainable accuracy,
let us assume that each synchronous frequency can be obtained to
a relative accuracy of $10^{-4}$, or to about .5~MHz.

As for systematic frequency errors we note from
Fig.~\ref{fivarfbar_3p4} that we are especially
sensitive to changes in average frequency.
For example, to double $S_{rms0}$ from its minimum, requires
a relative frequency change of only $4\times10^{-4}$.
If each cell frequency has an accuracy of $10^{-4}$, and there are
about 100 cells, the accuracy in the centroid 
frequency should be $\sim10^{-5}$. Therefore, the effect of this
type of systematic error should be negligible.

\begin{figure}[htb]
\centering
\epsfig{file=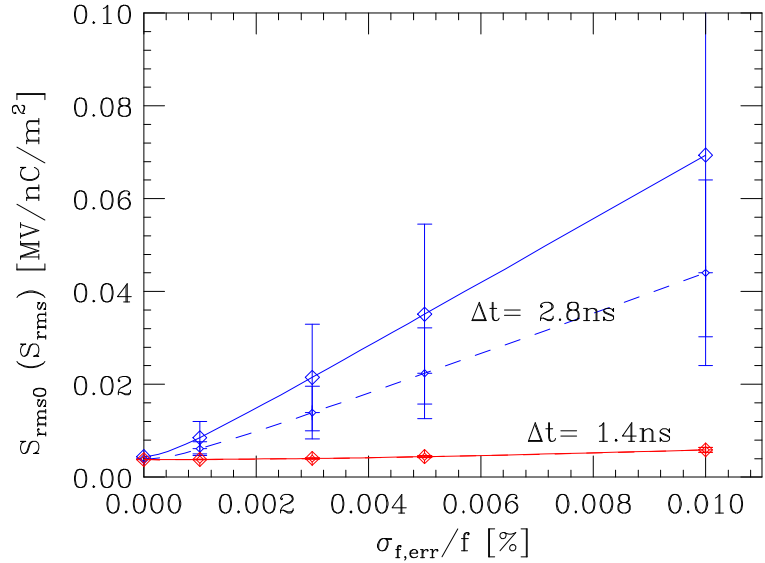, width=10cm}
\caption{
The optimal $3\pi/4$ structure: the effect of random frequency errors.
Shown are the relative (synchronous) frequency error, 
$\sigma_{f,err}$ {\it vs} $S_{rms0}$.
The dashed curves give $S_{rms}$. Each plotting symbol, with its error bars,
represents 400 different seeds.
}
\label{fierrs}
\end{figure}

As for random manufacturing errors, let us distinguish two types:
``systematic random'' and ``purely random'' errors. By ``systematic random'' we
mean errors, random in one structure, that are repeated in all structures
of the prelinac subsystem. ``Purely random''
 errors are, in addition, random from structure
to structure. In Fig.~\ref{fierrs} we give the resulting $S_{rms0}$
and $S_{rms}$,
for both bunch train configurations,
 when a
random error component is added to the (input) synchronous frequencies
of the optimal distribution.
With a frequency spacing of $\sim8\times10^{-4}$, an rms frequency error
of $10^{-4}$ is a relatively small perturbation. We see that 
for the alternate (1.4~ns) bunch spacing the effect of such a perturbation
is indeed very small, whereas for the nominal (2.8~ns) bunch spacing
the effect is large. The reason is that with the 1.4~ns bunch spacing
the beam sits near a half-integer resonance, whereas for the 2.8~ns
spacing it sits near the integer resonance. 
(Resonant multi-bunch wakefield effects are discussed
in Appendix~C.)
Note, however, that if we consider
the case of ``purely random'' machining errors, with a relative accuracy 
in synchronous frequencies of $10^{-4}$, and considering we have 
$N_{struc}=140$, 127, 41 structures in, respectively,
the prelinac, the $e^+$ drive linac, and the $e^-$
booster, then, with a $1/\sqrt{N_{struc}}$ reduction in sensitivity,
the appropriate abscissas in the figure become
 .8, .9, and $1.6\times10^{-5}$. At these points, for the 2.8~ns spacing, 
we see that $S_{rms0}$ is
only a factor $2\pm1$, $2\pm1$, $3\pm2$ times larger than the zero error result. 
Finally, as for the ``systematic random'' errors, it is difficult to judge
how large they might be in the real structure; however, they are likely
an order of magnitude less than the purely random errors, and should
therefore not yield a sum wake much larger than that due to the purely
random manufacturing errors.

If we make a weak damping, approximate calculation, by redoing the calculation
but now with $Q=1000$, we find no appreciable effect on the resonance behavior
for the $\Delta t=2.8$~ns case with frequency errors. For strong damping, taking
$Q=100$, however, we do find a suppression of the resonance effect.

\subsection{$e^+$ Booster}

For the $e^+$ booster (the L--band machine) each accelerator structure consists of
72 cells and the synchronous phase advance is taken to be $2\pi/3$.
The synchronous dipole mode distribution is taken to
be uniform, with $\bar f=2.05$~GHz and $\Delta_{\delta f}=3\%$.
We take $Q=20,000$. 
Note that in this case $\bar{f}\Delta t=2.87$; for the second bunch to 
sit on the zero crossing of the wakefield would require a shift in
frequency of $+4.5\%$ or $-13\%$.
For the L-band structure not much can be gained by changing the 
frequency spectrum.
We do gain, however, a factor of 8 reduction in wake in going from
S- to L-band, which, as we shall see, suffices.

\section{Tolerances}

For our designed structures we perform the tolerance calculations presented
in Section~2. 
To check the analytical estimates,
tracking, using the computer program LIAR\cite{LIAR}, was also performed.
 The numerical simulations
were simplified in that one macro-particle was used to represent each bunch, and
the bunch train was taken to be mono--energetic.
The analytical BBU results and those of LIAR, at 
the end of the four injector linacs,
are compared in Table~\ref{tasummary}.
 Results are given 
for bunch spacings of 1.4 and 2.8~ns. Under the heading ``Analytical'' are
given the rms of the sum wake $S_{rms}$, the rms of the growth factor $\Upsilon_{rms}$,
the maximum (within the bunch train) of the growth factor $\hat\Upsilon$, 
and the tolerance $r_t$ for $\delta_{\epsilon t}=10\%$, 
{\it i.e.} that ratio $y_0/\sigma_{y0}$ that results in 10\% emittance growth.
Under the heading ``Numerical'' we give the LIAR results: 
the maximum (within the bunch train) 
growth in normalized phase space $\hat\xi$ and the tolerance $r_t$
for 10\% emittance growth, both referenced to the centroid of the first bunch in the train.

\begin{table}[htb]
\caption{Beam break-up calculations for the two bunch spacings.
 Given are the 
rms of the sum wake $S_{rms0}$
(in units of MV/nC/m$^2$), the rms and the peak of the strength parameter,
 $\Upsilon_{rms0}$ and 
$\hat\Upsilon$, respectively, and the analytically obtained
tolerance $r_t$ for $\delta_{\epsilon t}=10\%$,
 {\it i.e.} that ratio $y_0/\sigma_{y0}$ that results in 10\% emittance growth. Also shown are 
LIAR results: the
peak blow-up in normalized phase space, $\hat\xi$, and the tolerance $r_t$.}
\label{tasummary}
\vskip6mm
\centering
\begin{tabular}{||c|l||c|c|c|c||c|c||} \hline \hline
     &      & \multicolumn{4}{c||}{Analytical}  & \multicolumn{2}{c||}{Numerical} \\ \cline{3-8}
\raisebox{1.5ex}[0pt]{$\Delta t$} &\raisebox{1.5ex}[0pt]{$\quad$Name}    &  $S_{rms0}$ & $\Upsilon_{rms0}$ & $\hat\Upsilon\ $ & \multicolumn{1}{c||}{$r_t$}& $\hat\xi\ $  & \multicolumn{1}{c||}{$r_t$}  \\ \hline\hline
&Prelinac &       .004        & .007 & .025 &  70                       &  .020  &  85        \\ \cline{2-8}
&$e^+$ Drive &    .004       &  .017 &  .066 &   25                      & .047   & 35         \\ \cline{2-8}
\raisebox{1.5ex}[0pt]{2.8ns}&$e^-$ Booster 
          &       .004       & .009 & .036 &   50                      &  .026  &   65       \\ \cline{2-8}
&$e^+$ Booster &  .12         & .119 & .227 &  3.8                      & .153   & 5.5        \\ \hline\hline
&Prelinac &       .004        & .004 & .019 &  115                      & .015   &  140       \\ \cline{2-8}
&$e^+$ Drive &    .004       &  .011 &  .049 &   45                      & .035   & 60         \\ \cline{2-8}
\raisebox{1.5ex}[0pt]{1.4ns}&$e^-$ Booster 
          & .004             & .006 & .027 &   80                      &   .019 &  105       \\ \cline{2-8}
&$e^+$ Booster &   .30        &  .205 &  .379 &  2.2                       & .257   & 3.0        \\ \hline\hline
\end{tabular}
\end {table}

\begin{table}[htb]
\caption{Effect of structure misalignments for the two bunch spacings.
 Given are the rms of the sum wake $S_{rms}$
(in units of MV/nC/m$^2$) and the tolerance for structure misalignments
for a 10\% emittance growth, $x_{at}$, both obtained analytically 
(Eqs.~\ref{eqmisa},\ref{eqxat}) and by LIAR. 
}
\label{tasummary2}
\vskip6mm
\centering
\begin{tabular}{||c|l||c|c||c||} \hline \hline
     &      & \multicolumn{2}{c||}{Analytical}  & \multicolumn{1}{c||}{Numerical} \\ \cline{3-5}
\raisebox{1.5ex}[0pt]{$\Delta t$} &\raisebox{1.5ex}[0pt]{$\quad$Name}    &  $S_{rms}$ &  \multicolumn{1}{c||}{$x_{at}[mm]$}&  \multicolumn{1}{c||}{$x_{at}$[mm]}  \\ \hline\hline
&Prelinac &       .004        &  2.9                       &     3.2          \\ \cline{2-5}
&$e^+$ Drive &    .004       &    100.                     &     120.         \\ \cline{2-5}
\raisebox{1.5ex}[0pt]{2.8ns}&$e^-$ Booster 
          &       .004       &   140.                    &      170.       \\ \cline{2-5}
&$e^+$ Booster &  .022         &   590.                     & \multicolumn{1}{c||}{}      \\ \hline\hline
&Prelinac &       .004        &   4.6                      &    4.8           \\ \cline{2-5}
&$e^+$ Drive &    .004       &     150.                     &   180.          \\ \cline{2-5}
\raisebox{1.5ex}[0pt]{1.4ns}&$e^-$ Booster 
          & .004             &  210.                     &      260.       \\ \cline{2-5}
&$e^+$ Booster &   .040        &    450.                      & \multicolumn{1}{c||}{}   \\ \hline\hline
\end{tabular}
\end {table}

We note from Table~\ref{tasummary} that for the S-band machines, 
$\Upsilon_{rms0}$ and $\hat\Upsilon$ are both small compared to 1, and that
the injection jitter tolerance for 10\% emittance growth is very large. For the L-band
machine, the $e^+$ booster, the tolerances are tighter but still acceptable.
We note also that the analytical $\hat\Upsilon$ agrees well with the numerical
$\hat\xi$, as do the two versions of $r_t$.

In Table \ref{tasummary2} we present misalignment results. Given are $S_{rms}$ and
the tolerance for structure misalignments, $x_{at}$, 
both as given analytically and by LIAR.
As discussed before, the meaning of $x_{at}$ is 
the rms misalignment that (for
an ensemble of machines) results, on average,
in a final emittance growth equal to a tolerance $\delta\epsilon_t$, which
in this case we set to 10\%. 
From Table~\ref{tasummary2} we see that the analytical and numerical results
agree well, and that the misalignment 
tolerances are all very loose. The tightest
tolerance is for the prelinac with nominal (2.8~ns)
bunch spacing, where the tolerance is still
an acceptable 3~mm. 

The effect of machining errors will tighten these tolerances for the
S-band machines with the nominal (2.8~ns) bunch spacing, due to the beam being
near the integer resonance. If machining adds a purely random error component that
is equivalent
to $10^{-4}$ frequency error, we saw earlier that (for the 2.8~ns bunch spacing case only)
this will tighten the injection jitter tolerances by about a factor $2\pm1$ for the prelinac and
$e^+$ drive linac, and about a factor of $3\pm2$ for the $e^-$ booster.
But even with this, the tolerances are still very loose.
The misalignment tolerances are affected less by machining errors. The prelinac
tolerance, with 2.8~ns bunch spacing, will become $\sim2\pm1$~mm;
for 95\% confidence in achieving $\delta_{\epsilon}=10\%$, the tolerance
becomes $\sim1\pm.5$~mm.

Finally, what is the random, 
cell-to-cell misalignment tolerance? Performing the
perturbation calculation described in Appendix~B, and calculating for 1000 different
random structures,
we find that $S_{rms}=.27\pm.12$~MV/nC/m$^2$ for $\Delta t=2.8$~ns, and 
$S_{rms}=.032\pm.003$~MV/nC/m$^2$ for $\Delta t=1.4$~ns.
We again see the effect of the integer resonance on the 2.8~ns option result. 
(To verify that this is the case,
we performed one run but with
the bunch spacing changed so that the beam sits
near the next half-integer resonance (11.5);
the result was that $S_{rms}$ dropped by a factor of 6.)
For the prelinac the cell-to-cell misalignment tolerance becomes
40~$\mu$m for the nominal (2.8~ns) bunch configuration and
600~$\mu$m for the alternate (1.4~ns) configuration.

\section{Conclusion}

We have demonstrated that by using detuning alone, the four injector
linacs can be built to sufficiently suppress the multi-bunch wakefield effects,
for both the nominal (2.8~ns) and alternate (1.4~ns) bunch spacings.
We have studied the sensitivity to multi-bunch beam break-up (BBU)
and to structure misalignments 
through analytical estimates and numerical tracking,
and shown that the tolerances to injection jitter, in the former case, and to
structure misalignments, in the latter case, are not difficult to achieve.
We have also studied the effect of manufacturing errors on these tolerances, and
have shown that if the errors are purely random, with an equivalent rms frequency error of
$10^{-4}$, then the other tolerances are still acceptable.
Finally, we have shown that the cell-to-cell misalignment tolerance is
$\gtrsim40$~$\mu$m.
 
For the L--band machine---the $e^+$ Booster---we have
shown that a uniform detuning of the dipole modes, 
with central frequency ${\bar f}=2.05$~GHz
and a total frequency
spread $\Delta_{\delta f}=3\%$, suffices.
For the S--band linacs---the Prelinac, the $e^+$ Drive Linac, and the $e^-$ 
Booster---we have shown that the 1.4~ns bunch spacing option forces us
to reduce the central frequency by 2.3\% from the nominal 4.012~GHz.
Doing this we lose 7\% in effective gradient, which, however, can be regained
by increasing the phase advance per cell from $2\pi/3$ to $3\pi/4$.
Our final, optimized distribution is trapezoidal in shape with
${\bar f}=3.920$~GHz, $\Delta_{\delta f}=5.8$\%, and tilt parameter
$\alpha= -0.2$. 

We have demonstrated in this report that the integer resonance, which we cannot
avoid given the two bunch train alternatives, can make us more sensitive to manufacturing
errors. Also, we have shown that the analytical, single-bunch beam break-up theory,
when slightly modified, can be useful in predicting the behavior of multi-bunch
beam break-up also.
 Given the rather loose tolerances demonstrated here makes us think 
that the S-band machines can be replaced with C-band ones that still
have reasonable tolerances, an option which may result in savings in cost,
though this needs further study.
Finally, we should reiterate that in this report we were
 concerned with the effects of
the modes in the first dipole passband only. 
With the wakefield of the first band
modes greatly suppressed by detuning, the effects
of the higher bands may no longer be insignificant. This problem will need to
be addressed in the future.

\section*{Acknowledgments}

The author thanks the regular attendees of the Tuesday JLC/NLC linac meetings
at SLAC for helpful comments and discussions on this topic,
and in particular T.~Raubenheimer, our leader, and V.~Dolgashev for
carefully reading parts of this manuscript.

\section*{Appendix A:\hfill\break Analytical Formula for Weak Multi-Bunch BBU}

\renewcommand{\theequation}{\mbox{A\arabic{equation}}}
\setcounter{equation}{0}

In Ref.~\cite{chao} 
an analytical formula for {\it single-bunch} beam break-up
in a smooth focusing linac, for the case without energy spread in
the beam,  is derived,
the so-called Chao-Richter-Yao (CRY) model for beam break-up. 
Suppose the beam is initially offset 
from the accelerator axis.
The beam break-up downstream is characterized by a strength parameter
$\Upsilon(t,s)$, where $t$ represents position within the bunch, and $s$
position along the linac.
When $\Upsilon(t,s)$ is small compared to 1, 
the growth in betatron amplitude in the linac is proportional to this parameter.
When applied to the special case of a uniform longitudinal
charge distribution, and a linearly growing wakefield, the result 
of the calculation becomes
especially simple. 
In this case the growth in orbit amplitude is given as an asymptotic
power series in $\Upsilon(t,s)$, and the series can
be summed to give a closed form, asymptotic
solution for single-bunch BBU.
The derivation of an analytic formula for {\it multi-bunch} BBU is almost 
a trivial modification of the CRY formalism.
We will here reproduce the important features of the 
single-bunch derivation of Ref.~\cite{chao}
(with slightly modified notation), and then
show how it can be modified to obtain a result applicable to multi-bunch BBU.
Note that we are interested in 
estimating the effect of relatively
weak multi-bunch BBU, caused by the somewhat complicated
wakefields of detuned structures.
The more studied multi-bunch BBU
problem, {\it i.e.} the effect on a bunch train of a single strong
mode, the so-called ``cumulative beam break-up instability''
(see, {\it e.g.} Ref.~\cite{Lau}), is a 
somewhat different problem to which our results are not meant to apply.

Let us consider the case of single-bunch beam break-up, where a beam is
initially offset by distance $y_0$ in a linac with 
acceleration and smooth focusing. 
We assume that there is no energy spread
within the beam.
The equation of motion is
\begin{equation}
{1\over E(s)}{d\over ds}
\left[E(s){dy(t,s)\over ds}\right]+ {y(t,s)\over\beta^2(s)}=
{e^2N_t\over E(s)}\int_{-\infty}^t dt^\prime\,y(t^\prime,s)\lambda_t(t^\prime)
W(t-t^\prime)\ ,\label{eqmotion}
\end{equation}
with $y(t,s)$ the bunch offset, a function of position within the bunch $t$,
and position along the linac $s$; with $E$ the beam energy, $[1/\beta(s)]$ the
betatron wave number, $eN_t$ the total bunch charge, $\lambda_t(t)$
the longitudinal charge distribution, and $W(t)$ the short-range 
dipole wakefield.
Our convention is that negative values of $t$ are toward the front of the 
bunch.
Let us, for the moment, limit ourselves to the problem of no acceleration and
$\beta$ a constant. 
A.~Chao in Ref.~\cite{chao} expands the solution to the equation of motion
for this problem in a perturbation series
\begin{equation}
y(t,s)=\sum_{n=0}^\infty y^{(n)}(t,s)\quad,\label{eqypert0}
\end{equation}
with the first term given by free betatron
oscillation [$y^{(0)}=y_0\cos (s/\beta)$].
He then shows that the solution for the higher terms
at position $s=L$, after many
betatron oscillations, is given by
\begin{equation}
y^{(n)}(t,L)\approx {y_0\over n!}\left(
{ie^2N_tL\beta\over 2E}
\right)^n R_n(t)e^{iL/\beta}\quad,\label{eqypert}
\end{equation}
with
\begin{eqnarray}
R_n(t)&=&\int_{-\infty}^t dt_1\,\lambda(t_1)W(t-t_1)
       \int_{-\infty}^{t_1} dt_2\,\lambda(t_2)W(t_1-t_2)\nonumber\\
      & &\cdots 
       \int_{-\infty}^{t_{n-1}} dt_n\,\lambda(t_n)W(t_{n-1}-t_n)\ ,
\label{eqrn}
\end{eqnarray}
and $R_0(z)=1$.
An observable $y$ is meant to be the real part of Eq.~\ref{eqypert0}.
The effects of adiabatic acceleration,
{\it i.e.} sufficiently slow acceleration so that the energy doubling
distance is large compared to the betatron wave length,
and $\beta$ not constant, can be added 
by simply replacing 
$(\beta/E)$ in Eq.~\ref{eqypert}
by $\langle\beta/E\rangle$, where
angle brackets indicate averaging along the
linac from $s=0$ to $s=L$.\footnote{Note that the terms $y_0 e^{iL/\beta}$
in Eq.~\ref{eqypert}, related to free betatron oscillation, 
also need to be modified in well-known ways
to reflect the dependence of $\beta$ on $E$.
It is the other terms, however,
which characterize BBU, in which we are interested.
}
For example, if the lattice is such that $\beta\sim E^\zeta$ then
$\langle\beta/E\rangle=(\beta_0/E_0)g(E_f/E_0,\zeta)$,
 where subscripts ``0'' and
``$f$'' signify, respectively, initial and final parameters, and
\begin{equation}
g(x,\zeta)= {1\over\zeta}\left({x^\zeta-1\over x-1}\right)
\quad\quad\quad[{\beta\sim E^\zeta}].
\end{equation}

Chao then shows that for certain simple combinations of bunch 
shape and wake function shape
the integrals in Eq.~\ref{eqrn} can
be performed analytically, and the result becomes  
an asymptotic series in powers of
a strength parameter.
For example, for the case of a uniform charge distribution of 
length $\ell$ (with the front of the bunch at $t=0$), and a wake
that varies as $W=W^\prime t$, the strength parameter is
\begin{equation}
\Upsilon(t,L)={e^2N_t LW^\prime t^2\beta_0\over 2E_0\ell}g(E_f/E_0,\zeta)
\quad.
\end{equation}
If $\Upsilon$ is small compared to 1, the growth is well approximated by 
$\Upsilon$.
 If $\Upsilon$ is large,
the sum over all terms can be performed to give
a closed form, asymptotic expression.

For {\it multi-bunch} BBU we are mainly concerned with the interaction of the
different bunches in the train, and will ignore wakefield forces
within bunches. The derivation is nearly identical to 
that for the single-bunch BBU.
However,
in the equation of motion, Eq.~\ref{eqmotion}, the independent variable $t$
is no longer a continuous variable,
but rather $t$ takes on discrete
values $t_m=m\Delta t$, where $m$ is a bunch index and $\Delta t$ is
the bunch spacing.
Also, $W$ now represents the long-range wakefield.
Let us assume that there are $M$, equally populated bunches in a train;
{\it i.e.} $N_t=MN$, with $N$ the particles per bunch.
The solution is again expanded in a perturbation series. In the solution,
Eq.~\ref{eqypert}, the $R_n(t)$, which are smooth functions
of $t$, are replaced by 
\begin{equation}
{\cal R}_m^{(n)}= {1\over M}\sum_{j=1}^{m-1} W[(m-j)\Delta t]{\cal R}_j^{(n-1)}
\quad,\label{eqcalr}
\end{equation}
(with ${\cal R}_j^0=1$), which is a function of a discrete 
parameter, the bunch index $m$. 
Note that ${\cal R}_m^{(1)}=S_m/M$, with $S_m$ the sum wake.

Generally the sums in Eq.~\ref{eqcalr} cannot be 
given in closed form, and therefore a closed, asymptotic expression
for multi-bunch BBU cannot be given. We can still, however, numerically
compute the individual terms equivalent to Eq.~\ref{eqypert} for the
single bunch case. For example, the first order term 
in amplitude growth is given by
\begin{equation}
\Upsilon_m= {e^2NLS_{m}\beta_0\over 2E_0} 
g(E_f/E_0,\zeta)\quad\quad\quad[m=1,\ldots,M]\ .
\end{equation}
If this term is small compared to 1 for all $m$, then BBU is well characterized
by $\Upsilon$. If it is not small, though not extremely
large, the next higher terms can be computed and their contribution added.
For $\Upsilon$ very large, this approach may not be very useful. 

From our derivation
we see that there is nothing that fundamentally distinguishes
our BBU solution from a single-bunch BBU solution. 
If we consider again the single-bunch calculation, for 
the case of a uniform charge distribution of length $\ell$, 
we see that we need to perform the integrations for $R_n$ in Eq.~\ref{eqrn}.
If we do the integrations numerically,
by dividing the integrals into discrete steps 
$t_n=(n-1)\Delta t$ and then performing
quadrature by rectangular rule, we end up with Eq.~\ref{eqcalr} with
$M=\ell/\Delta t$. The solution is the same as our multi-bunch solution. 
What distinguishes the multi-bunch from the single-bunch problem is that
the wakefield for the multi-bunch case is not normally
monotonic and does not vary smoothly with longitudinal position.
For such a case it may be more difficult to decide how many terms
are needed for the sum to converge. 

\begin{figure}[htb]
\centering
\epsfig{file=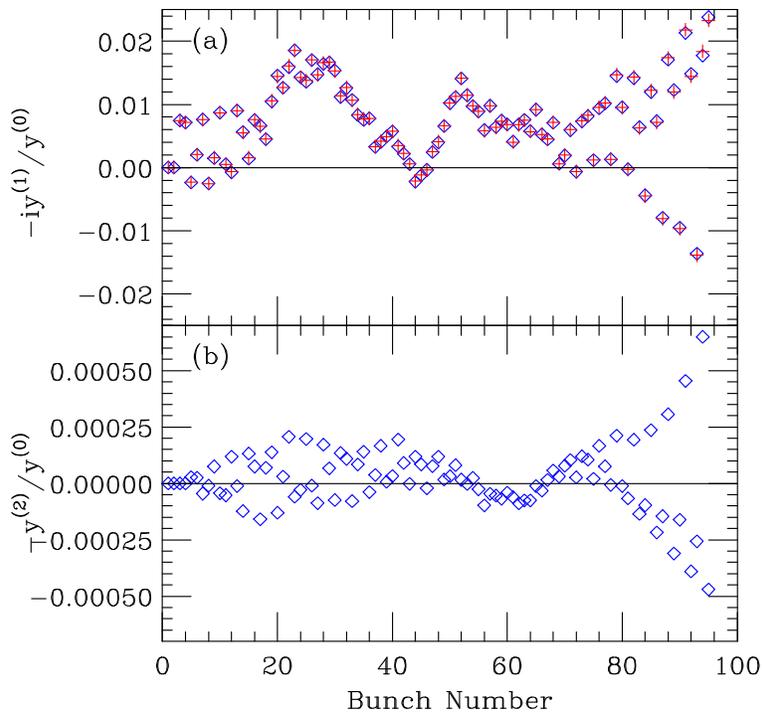, width=10cm}
\caption{
A numerical example: the NLC prelinac with the
optimized $3\pi/4$ S-band structure, but with $10^{-5}$ systematic
frequency errors, 
with the nominal (2.8~ns) bunch spacing (see the main text). 
$S_{rms}=.005$~MV/nC/m$^2$.
The diamonds give the first order~(a) and the second order~(b) perturbation
terms. The crosses in (a) give smooth focusing simulation results
with the free betatron term removed.
}
\label{fiperturb}
\end{figure}

In Fig.~\ref{fiperturb} we give a numerical example:
the NLC prelinac with the optimized $3\pi/4$ S-band structure,
but with $10^{-5}$ systematic
frequency errors, 
with the nominal (2.8~ns) bunch spacing (see the main text).
The diamonds give the first order~(a) and the second order~(b) perturbation
terms. The crosses in (a) give the results of a smooth focusing
simulation program (taking $\beta\sim E^{1/2}$),
where the free betatron term has been removed.
We see that the agreement is very good; {\it i.e.} the first order term
is a good approximation to the simulation results. 
In (b) we note that the next order term is much smaller.
For this example we find that even if we increase the current by an order
of magnitude the 1st order term alone remains a good approximation.

\section*{Appendix B:\hfill\break The Wakefield Due to Cell-to-Cell Misalignments}

\renewcommand{\theequation}{\mbox{B\arabic{equation}}}
\setcounter{equation}{0}

\begin{figure}[htb]
\centering
\epsfig{file=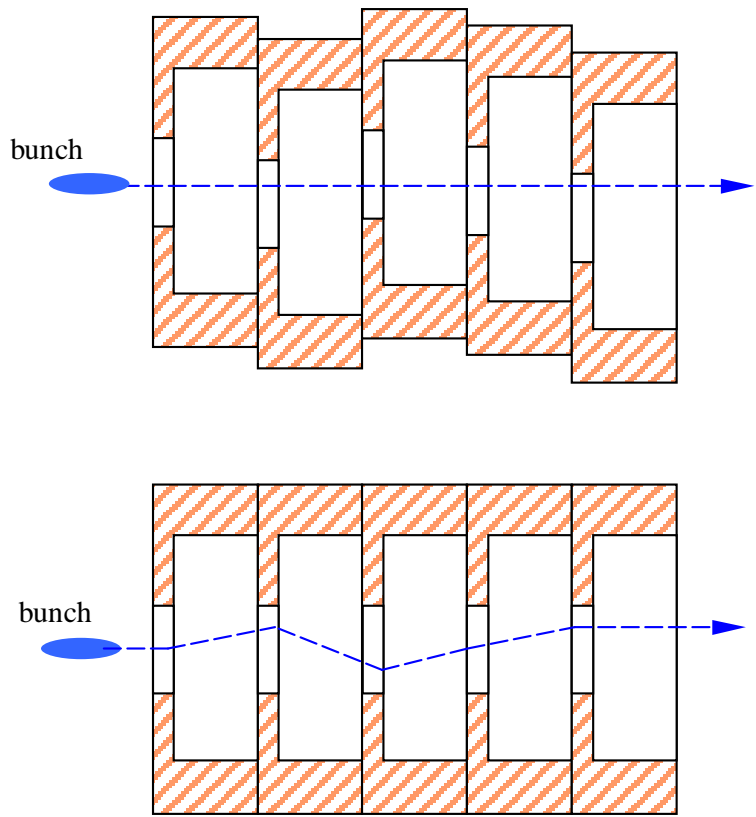, width=10cm}
\caption{
Sketches of part of a misaligned structure (top) and the model used
for the kick factor calculation (bottom). 
Note that the relative size of the misalignments here is much exaggerated.
}
\label{fistruct}
\end{figure}

We assume a structure is composed of many cells that are misaligned 
transversely by
amounts that are very small compared to the cell dimensions.
For such a case we assume that the mode frequencies are the same as in
the ideal structure,
and only the mode kick factors are affected. 
To first order we assume that for each mode, the kick factor for the beam
on-axis in the imperfect structure is the same as for the case with the beam
following the negative of the misalignment path in the error-free structure.
In Fig.~\ref{fistruct} we sketch a portion of such a misaligned structure (top)
and the model used for the kick factor calculation (bottom).
The sketch is meant to 
represent a disk-loaded structure that has been built up from a collection of
cups. Note that the relative size of the 
misalignements is exaggerated from what is expected, in order
to more clearly show the principle.
Given this model,
the method of calculation of the kick factors 
can be derived using the so-called 
``Condon Method''\cite{condon},\cite{morton} (see also \cite{bane}). 
Note that this 
application to cell-to-cell misalignments in an accelerator structure
is presented in Ref.~\cite{perturb}. 
The results of this perturbation method have been shown to be consistent
with those using a 3-dimensional scattering matrix analysis\cite{Valery}.
We will only sketch the derivation below.

Consider a closed cavity with perfectly conducting walls.
For such a cavity the Condon method 
expands the vector and scalar potentials,
in the Coulomb gauge, as a sum over the empty cavity modes.
As function of position $\bf x$ $(x,y,z)$ and time $t$ 
the vector potential in the cavity is given as
\begin{equation}
{\bf A}({\bf x},t)=\sum_\lm q_\lm(t){\bf a}_\lm({\bf x})\quad,
\end{equation}
where
\begin{equation}
\nabla^2{\bf a}_\lm + {\omega_\lm^2\over c^2}{\bf a}_\lm=0\quad,
\end{equation}
with $\omega_\lm$ the frequency of mode $\lm$, and
${\bf a}_\lm\times{\bf{\hat n}}=0$
on the metallic surface (${\bf{\hat n}}$ is a unit vector normal
to the surface).
Using the Coulomb gauge implies that $\nabla\cdot{\bf a}_\lm=0$.
The $q_\lm$ are given by
\begin{equation}
\ddot{q}_\lm+\omega_\lm^2 q_\lm={1\over 2U_\lm}\int_{\cal{V}}d\cal{V}\,
  {\bf j}\cdot {\bf a}_\lm\quad,
\end{equation}
with the normalization
\begin{equation}
{\epsilon_0\over2}\int_{\cal{V}}d\cal{V}\,
 {\bf a}_{\lm^\prime}\cdot{\bf a}_\lm= U_\lm\delta_{\lm\lm^\prime}\quad,
\end{equation}
with ${\bf j}$ the current density. Note that
the integrations are performed over the volume of the cavity $\cal{V}$.

The scalar potential is given as
\begin{equation}
\Phi({\bf x},t)=\sum_\lm r_\lm(t)\phi_\lm({\bf x})\quad,
\end{equation}
where
\begin{equation}
\nabla^2\phi_\lm + {\Omega_\lm^2\over c^2}\phi_\lm=0\quad,
\end{equation}
with $\Omega_\lm$ the frequencies associated
with $\phi_\lm$, and with $\phi_\lm=0$ on the metallic surface.
The $r_\lm$ are given by
\begin{equation}
r_\lm={1\over 2T_\lm}\int_{\cal{V}}d\cal{V}\,\rho\phi_\lm\quad,
\end{equation}
with $\rho$ the charge distribution in the cavity.
Note that one fundamental difference between the behavior
of $A({\bf x},t)$ and $\Phi({\bf x},t)$ is that
when there are no charges in the cavity the vector potential
can still 
oscillate whereas the scalar potential must be identically equal to 0.

Let us consider an ultra-relativistic driving charge $Q$ that passes through
the cavity parallel to the $z$ axis, and (for simplicity) a test charge
following at a distance $s$ behind on the same path.
 Both enter the cavity at position $z=0$
and leave at position $z=L$.
The transverse wakefield at the test charge is then
\begin{eqnarray}
{\bf W}(s)&=&{1\over QLx_0}\int_0^L dz\,\left[c\nabla_\bot A_z-\nabla_\bot\Phi\right]_{t=(z+s)/c}
            \nonumber\\
          &=&{1\over QLx_0}\sum_\lm\int_0^L dz\,\left[cq_\lm\left({z+s\over c}\right)\nabla_\bot
          a_{\lm z}(z)\right.\nonumber\\
          & &\quad\quad\quad\quad\quad\quad\quad\quad\quad
          -r_\lm\left({z+s\over c}\right)\nabla_\bot\phi_\lm(z)\bigg]\ ,
\label{eqconda}
\end{eqnarray}
where the integrals are along the path of the particle trajectory.
The parameter $x_0$ is a parameter for transverse offset
(the transverse wake is usually given in units of V/C per longitudinal
meter per transverse meter);
for a cylindrically-symmetric structure it is usually taken to be the
offset, from the axis, of the driving bunch trajectory.
For $s>L$ we can drop the scalar potential term
(it must be zero when there is no charge in the cavity), and
 the result can be written in the form\cite{bane}
\begin{equation}
{\bf W}(s)=\sum_\lm{c\over 2U_\lm\omega_\lm Lx_0}\Im{\rm m}\left[
V_\lm^*\nabla_\bot V_\lm \,e^{i\omega_\lm s/c}\right]
\quad\quad[s>L]\ ,
\label{eqcondc}
\end{equation}
with 
\begin{equation}
V_\lm=\int_0^L dz\,a_{\lm z}(z)e^{i\omega_\lm z/c}\quad.
\end{equation}
Note that the arbitrary constants associated
with the parameter ${\bf a}_\lm$ in the numerator and the denominator of 
Eq.~\ref{eqcondc} cancel.
Note also that---to the same arbitrary constant---$|V_\lm|^2$
 is the square of the voltage lost by
the driving particle to mode $\lm$ and 
$U_\lm$ is the energy stored in mode $\lm$.

Consider now the case of 
a cylindrically-symmetric, multi-cell accelerating cavity, and let us limit 
our concern to the effect of the dipole modes of such a structure.
We will allow the charges to move on an arbitrary, zig-zag path in the $x-z$ plane
that is close to the axis, and for which the slope is everywhere small
(so that $\nabla_\bot\sim \partial/\partial x$).
For dipole modes in a cylindrically-symmetric, multi-cell
accelerator structure, it can shown that
the synchronous component of $a_{\lm z}$ (the only component that, on
average, is important) can be written in the form $a_{\lm z}=xf_\lm(z)$
(see {\it e.g.} Ref.~\cite{Trans}).
Then Eq.~\ref{eqcondc} becomes
\begin{eqnarray}
W_x(s)&=&\sum_\lm{c\over 2U_\lm\omega_\lm Lx_0}\times\label{eqconde}\\
& &\hspace*{-11mm}\times\Im{\rm m}\left[e^{i\omega_\lm s/c}
\int_0^L dz^\prime\, x(z^\prime)f_\lm(z^\prime)e^{-i\omega_\lm z^\prime/c}
\int_0^L dz\, f_\lm(z)e^{i\omega_\lm z/c}
\right]\ [s>L]\ .\nonumber
\end{eqnarray}
Note that this equation can be written in the form:
\begin{equation}
W_x(s)= \sum_\lm 2k^\prime_{\lm}\sin\left({\omega_\lm s\over c}+\theta_\lm\right) 
\quad\quad[s>L]\ ,
\label{eqcondb}
\end{equation}
with $k^\prime_{\lm}$ a kind of kick factor and $\theta_\lm$ the phase
of excitation of mode $\lm$.
Note that in the special case
where the particles move parallel to the axis, at offset $a$,
$k^\prime_{\lm}=k_\lm=c|V_\lm|^2/(4U_\lm\omega_\lm a^2L)$,
 the normal kick factors for the structure,
and $\theta_\lm=0$.
For this case it can be shown that Eq.~\ref{eqcondb} is valid for
all $s>0$\cite{bane}.
Finally, note that, for the general case,
Eq.~\ref{eqcondb} can obviously not be  extrapolated down to $s=0$,
since it implies that $W_x(0)\not=0$, which we believe is nonphysical,
implying that a particle can kick itself transversely.
To obtain the proper equation valid down to $s=0$ we
would need to include the scalar potential term that was dropped in going
from Eq.~\ref{eqconda} to Eq.~\ref{eqcondc}.  

Our derivation, presented here, is technically applicable only to structures
for which all modes are trapped. The modes will be trapped at least at the
ends of the structure, if the connecting beam tubes have sufficiently
small radii and the dipole modes do not couple to the fundamental
mode couplers in the end cells.
 For detuned structures, like those in the injector
linacs discussed in this report, most modes are trapped internally within
a structure, and those that do extend to the ends couple only weakly
to the beam; for such structures the results here can also be applied,
even if the conditions on the beam tube radii and the fundamental mode
coupler do not hold.
We believe that even for the damped, detuned structures of the
main linac of the JLC/NLC, which are similar, though they have
manifolds to add weak damping to the wakefield, a result very similar to that
presented here applies.

To estimate the wakefield associated with very small, random cell-to-cell
misalignments in accelerator structures we assume that 
we can use the mode eigenfrequencies and eigenvectors of the error-free
structure.
We obtain these from the circuit program.
Then to find the kick factors we replace
$x(z)$ in the first integral in Eq.~\ref{eqconde}
by the zig-zag path representing the negative of the cell misalignments, a path
we generate using a random number generator.
The normalization factor $x_0$ is set to the rms of the misalignments.
How can we justify using this method for finding the wake
at the spacing of the bunch positions?
For example, for
the $3\pi/4$ S-band structure, the alternate bunch spacing
is only 42~cm whereas the whole structure length
$L=4.46$~m. Therefore, in principle,
Eq.~\ref{eqconde} is not valid until the 11$^{\rm th}$ bunch spacing.
We believe, however, that the scalar potential fields will not
extend more than one or two cells 
behind the driving charge (the cell length is 4.375~cm), and
therefore this method will be a good approximation at all bunch
positions behind the driving charge.
This belief should be tested in the future by repeating
the calculation,
but now also including the contribution from scalar potential terms.

\begin{figure}[htb]
\centering
\epsfig{file=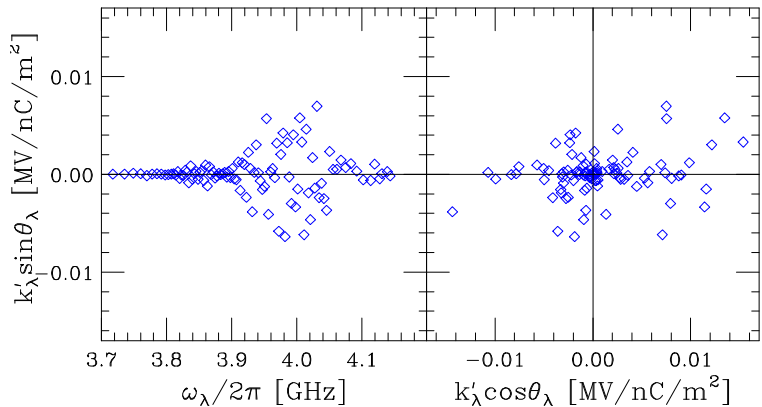, width=10cm}
\caption{
The kick factors and phases of the modes for a cell-to-cell
misalignment example. The structure is the optimized $3\pi/4$ S-band structure
(see the main text). For this example, for the nominal (2.8~ns) bunch spacing,
$S_{rms}=.32$~MV/nC/m$^2$.
}
\label{fiscatter}
\end{figure}

In Fig.~\ref{fiscatter} we give a numerical example. 
Shown, for the optimized $3\pi/4$ S-band structure (see the
main text),  
are the kick factors and the phases of the modes as calculated
by the method described in this section. Note that $\theta_\lambda$ is
not necessarily small.

\section*{Appendix C:\hfill\break Resonant Multi-Bunch Wakefield Effects}

\renewcommand{\theequation}{\mbox{C\arabic{equation}}}
\setcounter{equation}{0}

It is easy to understand
how resonances can arise in a linac with
bunch trains. 
Consider the case of the interaction
of the beam with one single structure mode.
The leading bunch enters the structure 
offset from the axis and excites the mode.
If the bunch train is sitting on an integer resonance,
{\it i.e.} if $f\Delta t=n$, with $f$ the mode frequency,
$\Delta t$ the bunch spacing, and $n$ an integer,
then when the 2nd bunch arrives it will excite the mode at the same phase
and also obtain a kick due to the wakefield of the first bunch.
The $m{\rm th}$ bunch will also excite the mode in the same phase
and obtain $(m-1)$ times the kick from the wakefield that the second bunch experienced
(for simplicity we assume the mode $Q$ is infinity).
On the half-integer resonance, {\it i.e.} when $f\Delta t=n+.5$, 
the $m{\rm th}$ bunch will also receive kicks 
from the wakefield left by the
earlier bunches, but in this case the kicks will
 alternate in direction, and no resonance builds up.
For a transverse wakefield effect, such as we are interested in, however,
this simple description of the resonant
interaction needs to be modified slightly.
For this case the wake varies as
$\sin(2\pi ft)$, and neither the integer nor
 the half-integer resonance condition 
will excite any wakefield for the following bunches. 
In this case resonant growth
is achieved at a slight deviation from 
the condition $f\Delta t=n$, as is shown below.

In the following, for simplicity,
 we will use the ``uncoupled'' model (which is described
in Chapter~3 of the main text) to 
investigate resonant effects in the sum wake for a structure with modes with a
uniform frequency distribution. 
The point of using the uncoupled model is that it allows us to
study the effect of an idealized, uniform frequency distribution.
As we have seen in the main text, an ideal (input) frequency distribution
becomes distorted by the cell-to-cell coupling of an accelerator structure.
As example we will use the
parameters of a simplified version (all kick factors are equal) of the
optimized $3\pi/4$ S-band structure described in
the main text; for bunch structure
we consider the nominal bunch spacing ($\Delta t=2.4$~ns).
The results for the real structure, with coupled modes, will be
slightly different yet qualitatively the same.
Note that we are also aware of a different analysis of
resonant multi-bunch wakefield effect\cite{schulte}.

\begin{figure}[htb]
\centering
\epsfig{file=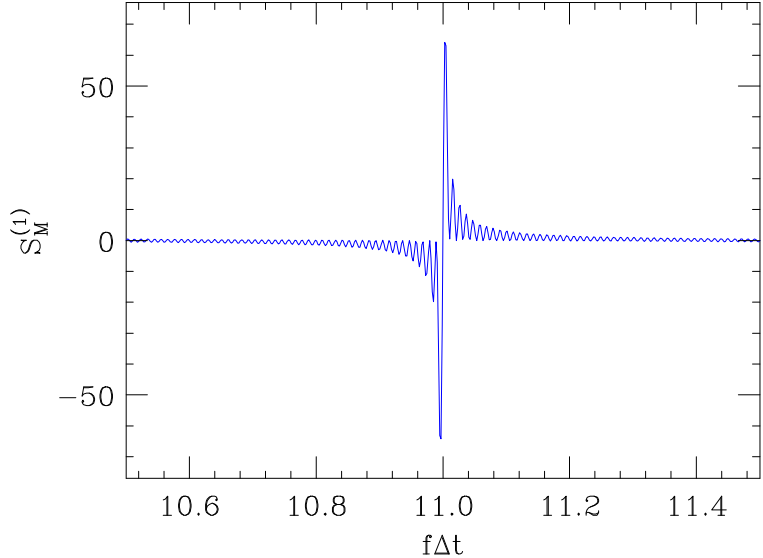, width=10cm}
\caption{
The sum wake at the last bunch in a train {\it vs} bunch spacing, due to a single mode
(Eq.~\ref{eqres1}); $M=95$ bunches.
}
\label{fires1}
\end{figure}

Consider first 
the case of a structure with only one dipole mode, with frequency $f$,
and a kick factor that we will normalize (for simplicity) to $1/2$.
Suppose there are $M$ bunches in the bunch train.
The sum wake at the $m{\rm th}$ bunch is given by
\begin{eqnarray}
S_m^{(1)}(f\Delta t)&=&\sum_{i=1}^m\sin\left(2\pi[i-1] f\Delta t\right)\nonumber\\
   &=& {\sin\left(\pi[m-1]f\Delta t\right)\sin\left(\pi mf\Delta t\right)
\over\sin\left(\pi f\Delta t\right)}\quad. 
\label{eqres1}
\end{eqnarray}
As with the nominal (2.8~ns) bunch spacing in the S-band prelinacs, let us,
for an example, consider $M=95$ bunches
and the region near the 11th harmonic. In Fig.~\ref{fires1}
 we plot $f\Delta t$ {\it vs}
the sum wake for the $M$th (the last) bunch, $S_M^{(1)}$, near the
11th integer resonance. 
It can be shown that, if $M$ is not small, the largest resonance peaks 
(the extrema of the curve) are at
\begin{equation}
f\Delta t\approx n\pm {3\over8M}\quad\quad[M\ {\rm not\ small}]\quad,
\end{equation}
with values $\pm.72M$.
Note that at the exact integer and half-integer resonant spacings the sum wake is zero.

\begin{figure}[htb]
\centering
\epsfig{file=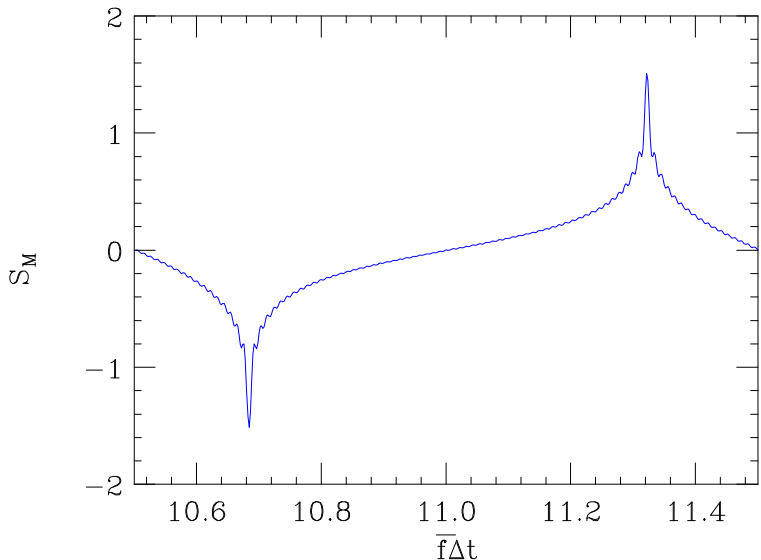, width=10cm}
\caption{
The sum wake at the last bunch in a train {\it vs} bunch spacing,
due to a uniform distribution of mode frequencies
(Eq.~\ref{equnib}).
 The total frequency spread $\Delta_{\delta f}=5.8\%$, and
$N_c=102$.
}
\label{fires2}
\end{figure}

Now let us consider a uniform distribution of mode frequencies.
For simplicity we will let all the kick factors be equal, and be normalized to
$1/2$. 
The sum wake, according to the uncoupled model, becomes
\begin{equation}
S_m({\bar f}\Delta t)={1\over N_c}\sum_{n=1}^{N_c}S_m^{(1)}\left[{\bar f}\Delta t\left(
1+{(n-N_c/2)\over N_c}\Delta_{\delta f}\right)\right]\quad,\label{equnib}
\end{equation}
with $N_c$ the number of cells (also the number of modes),
 $\bar f$ the central frequency, and $\Delta_{\delta f}$
the total (relative) width of the frequency distribution.
As an example, let us consider
the optimized $3\pi/4$ S-band structure, with $N_c=102$ and 
$\Delta_{\delta f}=5.8\%$. The sum wake at the last (the $M$th) 
bunch 
position, $S_M$, is plotted as function of ${\bar f}\Delta t$ in Fig.~\ref{fires2}.
Note that the uniform frequency distribution 
appears to suppress the integer resonance.
The extrema of the curve (the ``horns'') that are seen
at ${\bar f}\Delta t=11\pm.32$ 
are resonances due to the edges of the frequency distribution,
with the condition ${\bar f}\Delta t\approx11/(1\pm\Delta_{\delta f}/2)$.
Note, however, that the sizes of even these
spikes are small compared to those of the single mode case.

\begin{figure}[htb]
\centering
\epsfig{file=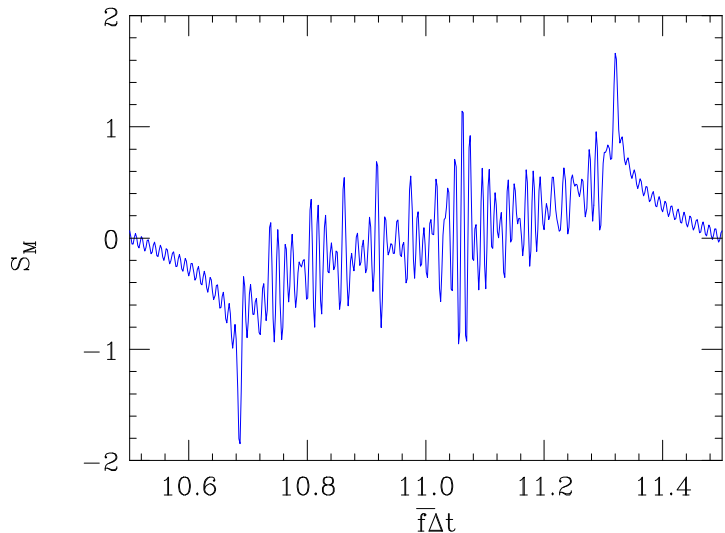, width=9.5cm}
\caption{
The sum wake at the last bunch in a train {\it vs} bunch spacing,
due to a uniform distribution of frequencies,
including frequency errors.
 The total frequency spread $\Delta_{\delta f}=5.8\%$, the number of
modes $N_c=102$, and rms relative frequency error is $10^{-4}$.
}
\label{fires3}
\end{figure}

\begin{figure}[htb]
\centering
\epsfig{file=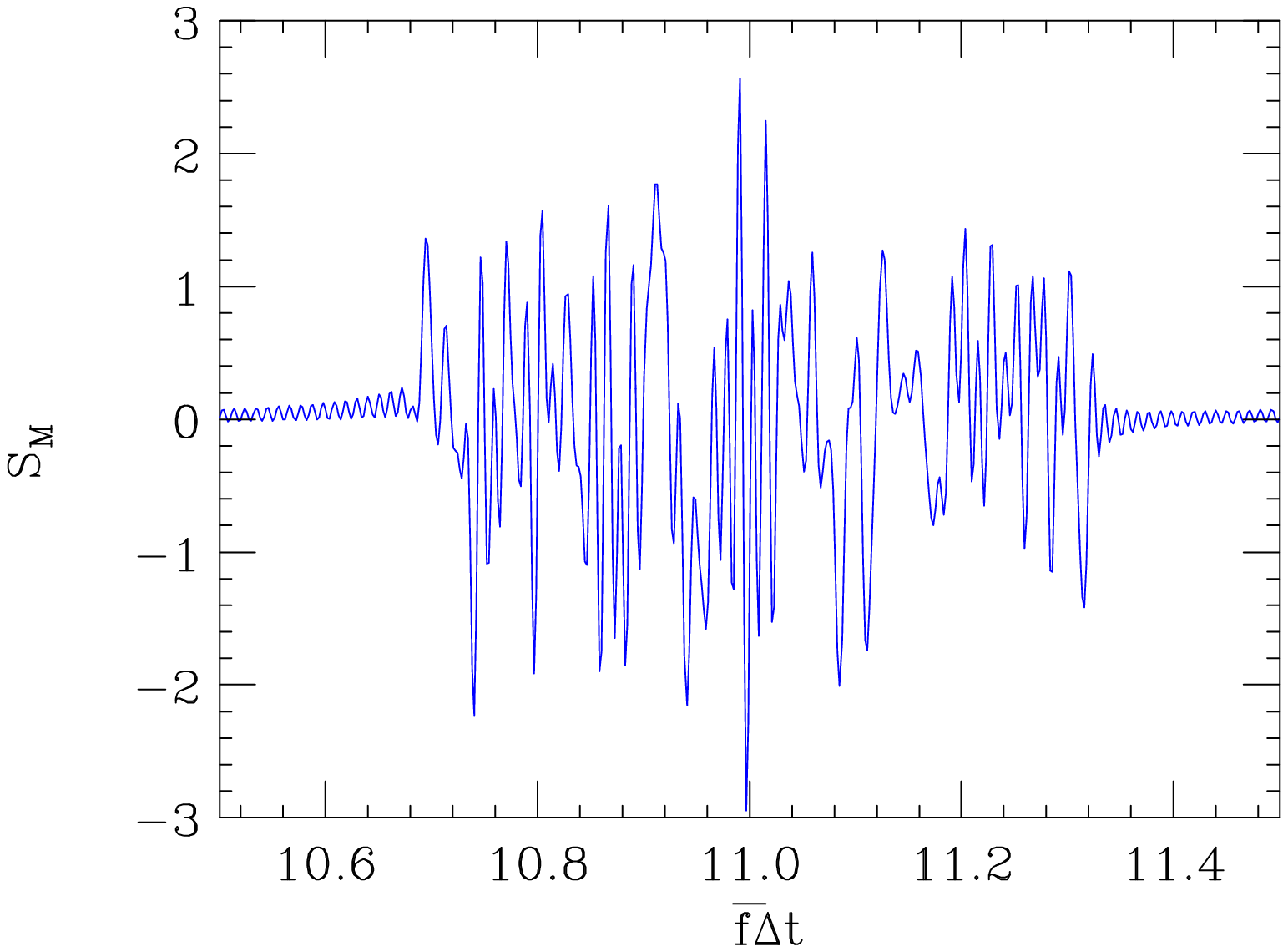, width=9.5cm}
\caption{
The sum wake at the last bunch in a train {\it vs} bunch spacing,
due to a uniform distribution of frequencies,
including random misalignment errors with rms 1.
The total frequency spread $\Delta_{\delta f}=5.8\%$ and then number of modes
$N_c=102$.
}
\label{fires4}
\end{figure}

Suppose we add frequency errors to our model. We can do this by, 
in each term in the sum of Eq.~\ref{equnib},
multiplying the frequency by the factor $(1+\delta f_{err}r_n)$,
 with $\delta f_{err}$ the rms (relative) frequency error and $r_n$ a random number
 with rms 1.
Doing this, considering a uniform distribution
in frequency errors with rms $\delta f_{err}=10^{-4}$,
 Fig.~\ref{fires2} becomes
Fig.~\ref{fires3}. 
Note that this 
perturbation is small compared to the frequency spacing
 $5.7\times10^{-4}$, so it does not really change the frequency distribution
significantly.
Nevertheless, because of 
resonance-like behavior 
we can see a large effect on $S_M$ throughout the range between the
horns of Fig.~\ref{fires2} ($10.68\leq {\bar f}\Delta t\leq 11.32$). 
To model cell-to-cell misalignments, we multiply
each term in the sum of Eq.~\ref{equnib} by the random factor $r_n$.
The results,
for a uniform distribution of 
errors with rms 1, are shown in Fig.~\ref{fires4}. Again resonance-like
behavior
is seen throughout the range between the horns of Fig.~\ref{fires2}. 

We can understand these results in the following manner:
Only when there are no errors 
does using a uniform frequency distribution
suppress the resonance in the region near the integer resonance.
But otherwise, using a uniform frequency distribution 
basically only reduces the size of the
resonances, at the expense of extending
the range in bunch spacings where they can be
excited. Instead of being localized in the region near the integer resonance
(${\bar f}\Delta t\approx n$), 
resonance-like behavior can now be excited anywhere between the limits
\begin{equation}
({\bar f}\Delta t)_\pm ={n\over1\mp\Delta_{\delta f}/2}\quad.
\end{equation}
Note that this implies that if $\Delta_{\delta f}>1/({\bar f}\Delta t)$,
then the resonance-like behavior 
cannot be avoided no matter what bunch spacing 
(fractional part) is chosen. For example, for the X-band linac in the NLC,
where the total width of the dipole frequency distribution 
(of the dominant first band modes) is 10\%,
even for the alternate (1.4~ns) bunch spacing, where the integer part
of ${\bar f}\Delta t$ is 21, the resonance region cannot be avoided.

\end{document}